\documentclass[aps,prb,twocolumn,amsmath,amssymb,hidelinks,superscriptaddress]{revtex4-2}

\usepackage{bm}
\usepackage{graphicx}
\usepackage{verbatim} 
\usepackage[colorlinks]{hyperref}
\hypersetup{
  colorlinks,
  citecolor=blue,
  linkcolor=blue,
  urlcolor=blue
}
\usepackage{orcidlink}

\renewcommand{\vec}[1]{{\bf #1}}

\begin{document}

\title{Quantum kinetic theory of the semiclassical side jump, skew scattering and longitudinal velocity}

\author{Da Ma}
\affiliation{College of Sciences, Northeastern University, Shenyang 110819, China}
\affiliation{Interdisciplinary Center for Theoretical Physics and Information Sciences, Fudan University, Shanghai 200433, China}
\affiliation{Hefei National Laboratory, Hefei 230088, China}
\author{Zhi-Fan Zhang\,\orcidlink{0000-0002-7063-4020}}
\affiliation{Interdisciplinary Center for Theoretical Physics and Information Sciences, Fudan University, Shanghai 200433, China}
\author{Hua Jiang\,\orcidlink{0000-0002-8626-735X}}
\thanks{Corresponding author. \\ \href{mailto:jianghuaphy@fudan.edu.cn}{jianghuaphy@fudan.edu.cn} }
\affiliation{Interdisciplinary Center for Theoretical Physics and Information Sciences, Fudan University, Shanghai 200433, China}
\affiliation{State Key Laboratory of Surface Physics, Fudan University, Shanghai 200433, China}
\author{X. C. Xie}
\affiliation{Interdisciplinary Center for Theoretical Physics and Information Sciences, Fudan University, Shanghai 200433, China}
\affiliation{Hefei National Laboratory, Hefei 230088, China}
\affiliation{International Center for Quantum Materials, School of Physics, Peking University, Beijing 100871, China}
\date{\today} 

\begin{abstract}
The semiclassical Boltzmann equation is widely used to study transport effects. However, being semiclassical and borrowing heavily from classical mechanics, the formalism calls for verification from the perspective of quantum mechanics. Although previous works discussed the relation between the quantum density matrix and the semiclassical formalism, direct comparison, especially of disorder effects, including side jumps and skew scattering in the two approaches, has not been fully conducted. In this work, we systematically and directly compare the semiclassical Boltzmann equation and its counterpart arising from the density matrix. We find that there is an additional correction to the side-jump velocity, the longitudinal velocity, which is longitudinal in the leading order, and its resultant current does not require time-reversal symmetry breaking. Moreover, we find the semiclassical side-jump collision integral is an approximation of the quantum result at moderate temperatures, and it also contains a correction induced by the longitudinal velocity. We also show that the scattering rate obtained from the density matrix agrees with the semiclassical results. Our work illuminates the quantum roots of the semiclassical Boltzmann equation.  
\end{abstract}
\maketitle

\section{Introduction}
The semiclassical Boltzmann equation has proven to be a simple and powerful tool to handle electric transport~\cite{Ziman1972,Cohen1973,Ashcroft1976,Sinitsyn2007,Nagaosa2010,DiXiao2010}. The semiclassical formalism combines the kinetic theory of classical gases and the quantum theory of solids, and it has been amended multiple times as the anomalous velocity induced by Berry curvature, skew scattering, side jumps, and other contributions have been introduced into it. One naturally wonders whether it is rigorous enough and whether it completely takes the main contribution into consideration. Previous studies tried to verify the semiclassical Boltzmann equation with the Kubo formula~\cite{Mahan2000,Sinitsyn2007PRB} and the Keldysh formalism~\cite{Mahan2000,Onoda2006,Konig2021}. For instance, the simple bubble diagram with the diagonal components of the velocity operator in the Kubo-Streda formula leads to the result of the Boltzmann equation with conventional group velocity, while the off-diagonal part of the velocity operator with the bubble diagram gives rise to the anomalous velocity induced by Berry curvature, and skew-scattering and side-jump effects can be accounted for if higher-order diagrams are introduced~\cite{Mahan2000,Sinitsyn2007PRB}. However, these approaches are generally convoluted and involve laborious calculations, and they are often opaque or model dependent, making direct comparison difficult.  

As the semiclassical Boltzmann equation and the quantum density matrix are closely intertwined, multiple attempts have been made to verify the semiclassical formalism with the density matrix~\cite{Kohn-Luttinger1957,Luttinger1958,PhysRevB.81.125332,Culcer2017,Sekine2017,Xiao2018,Atencia2022,PhysRevB.110.245406}. Yet classic works long predate the advent of modern concepts including Berry curvature and the modern development of the semiclassical transport theory, and recent papers have not fully resolved issues of impurity scattering. For example, are there other corrections to the velocity besides the side-jump velocity? How does the semiclassical side-jump collision integral square with the density matrix? Why do the density matrix approach and the semiclassical formalism disagree on the skew-scattering rate? 

In this work, we directly and systematically compare the semiclassical Boltzmann equation and the density matrix and address the questions above. We focus on spatially homogeneous systems in a uniform electric field. We show that there is a correction to the side-jump velocity, dubbed the longitudinal velocity, which is longitudinal in the leading order, and its resultant current does not require time-reversal symmetry breaking. We also show that the semiclassical side-jump collision integral is an approximation of  its quantum counterpart at moderate temperatures, and the collision integral also contains a correction originating from the longitudinal velocity. As for the skew-scattering rate, we fully perform averages over the impurity configuration ensemble and find that the rate at order $V^{3}$ and $V^{4}$ agrees with that obtained from other methods. Our work shows that the semiclassical Boltzmann formalism is more than a mishmash of the classical kinetics of gases and quantum mechanical treatment of wave packets and that it has roots in the density matrix, further illuminating the quantum mechanical background of the semiclassical Boltzmann equation.  

\section{The Boltzmann equation with a uniform electric field}
\label{sec:Boltzmann}
We consider spatially homogeneous systems in a uniform electric field and take the general form of the Hamiltonian, $\hat{H} = \hat{H}_{0} + \hat{U} + \hat{H}_{E}$, where $\hat{H}_{0}$ is the bare Hamiltonian without impurities, $\hat{U}$ describes impurity scattering, and $\hat{H}_{E} (t) = e \hat{ {\bf r} } \cdot {\boldsymbol{\mathcal{E}}}(t)$ embodies an applied electric field. Here, $\hat{ {\bf r} }$ is the position operator. Throughout this paper, we assume bands are nondegenerate. Moreover, we focus on the semiclassical regime, where the frequency of the applied electric field ${\boldsymbol{\mathcal{E}}}(t)$, $\omega$, is low, i.e., $\hbar \omega \ll \vert \varepsilon_{ {\bf k} }^{ m } - \varepsilon_{ {\bf k} }^{ m^{\prime} } \vert$ ($m \ne m^{\prime}$, and $\varepsilon_{ {\bf k} }^{ m }$ is the dispersion of the $m$th band), and real interband transition is ignored. We assume that bands are separated so that $\delta \left( \varepsilon_{ {\bf k} }^{ m } - \varepsilon_{ {\bf k}^{ \prime } }^{ m^{ \prime } } \right)$ implies $m = m^{ \prime }$.

We add a subscript I to operators to signify the interaction picture. For example, $\hat{A}_{I}(t) \equiv e^{ \frac{i}{\hbar} \hat{H}_{0} t} \hat{A} e^{ - \frac{i}{\hbar} \hat{H}_{0} t}$, where $\hat{A}$ is an operator in the Schr{\"o}dinger picture while $\hat{A}_{I}(t)$ is in the interaction picture. The interaction picture helps to eliminate the commutator $[\hat{H}_{0}, \hat{\rho}]$ in the quantum Liouville equation,
\begin{equation}
\label{eq:Liouville-1}
\frac{d}{dt} \hat{\rho}_{\mathrm{I}}(t) + \frac{i}{\hbar} \left[ \hat{H}_{E, I} (t), \hat{\rho}_{\mathrm{I}}(t)\right] + \frac{i}{\hbar} \left[ \hat{U}_{\mathrm{I}}(t), \hat{\rho}_{\mathrm{I}}(t)\right] = 0,
\end{equation}
where $\hat{\rho}$ is the density matrix. The interaction picture is crucial to directly connect the quantum Liouville equation to the semiclassical Boltzmann equation. 

We assume that the system reaches equilibrium at $t \to - \infty$, and the electric field is slowly turned on, ${\boldsymbol{\mathcal{E}}}(t) = 1/2 \left( \vec{E} e^{ i \left( \omega - i\eta \right) t} + \mathrm{c.c.} \right)$, $\eta \to 0^{+}$. We denote the density matrix at equilibrium as $\hat{\rho}_{0}$, which is diagonal in the band indices, $\langle m, {\bf k} \vert \hat{\rho}_{0} \vert m^{\prime}, {\bf k} \rangle = \delta_{m m^{\prime}} f_{0, {\bf k} }^{m}$, where $\vert m, {\bf k} \rangle$ is an eigenstate of $\hat{H}_{0}$ in the Bloch representation with band number $m$ and crystal momentum ${\bf k}$ in the Schr\"odinger picture, and $f_{0, {\bf k} }^{m}= 1/\left( e^{ \beta \left( \varepsilon_{ {\bf k} }^{ m } - \mu \right) } + 1 \right)$ is the Fermi-Dirac distribution function~\cite{Culcer2017}, where $\mu$ is the chemical potential, and $\beta = 1/\left( k_{ \mathrm{B} } T \right)$. For simplicity, we now consider only the response linear in ${\boldsymbol{\mathcal{E}}}$, and the density matrix $\hat{\rho}_{\mathrm{I}}(t)$ can be decomposed into the equilibrium part $\hat{\rho}_{0}$ and the linear-response part $\hat{\rho}_{E, \mathrm{I}}(t)$,
\begin{equation}
\hat{\rho}_{\mathrm{I}}(t) = \hat{\rho}_{0} + \hat{\rho}_{E, \mathrm{I}}(t).
\end{equation}
This allows us to keep the terms in Eq.~(\ref{eq:Liouville-1}) to the first order of ${\boldsymbol{\mathcal{E}}}$,
\begin{equation}
\label{eq:Liouville-2}
\frac{d}{dt} \hat{\rho}_{E, \mathrm{I}}(t) + \frac{i}{\hbar} \left[ \hat{H}_{E, I} (t), \hat{\rho}_{0} \right] + \frac{i}{\hbar} \left[ \hat{U}_{\mathrm{I}}(t), \hat{\rho}_{E, \mathrm{I}}(t)\right] = 0,
\end{equation}
Here we take the approximation $\left[ \hat{U}_{\mathrm{I}}(t), \hat{\rho}_{0} \right] = 0$, namely, we assume that impurity scattering is weak and the density of states is not affected by disorder. 

Equation~(\ref{eq:Liouville-2}) can be rewritten in the integral form as
\begin{align}
\label{eq:rho-integral}
\hat{\rho}_{E, \mathrm{I}}(t) = & - \frac{i}{\hbar} \int_{-\infty}^{t} d t_{1} \left[ \hat{H}_{E, I} (t_{1}), \hat{\rho}_{0} \right] \nonumber \\ 
& - \frac{i}{\hbar} \int_{-\infty}^{t} d t_{1} \left[ \hat{U}_{\mathrm{I}}(t_{1}), \hat{\rho}_{E, \mathrm{I}}(t_{1})\right].
\end{align}
We plug Eq.~(\ref{eq:rho-integral}) into the last commutator on the left hand side of Eq.~(\ref{eq:Liouville-2}),
\begin{align}
\label{eq:Liouville-3}
&\frac{d}{dt} \hat{\rho}_{E, \mathrm{I}}(t) + \frac{i}{\hbar} 
\left[ \hat{H}_{E, I} (t), \hat{\rho}_{0} \right] \nonumber \\
+ & \frac{1}{\hbar^{2}} \int_{-\infty}^{t} d t_{1} \left[ \hat{U}_{\mathrm{I}}(t), \left[ \hat{H}_{E, I} (t_{1}), \hat{\rho}_{0} \right] \right] \nonumber \\
+ & \frac{1}{\hbar^{2}} \int_{-\infty}^{t} d t_{1} \left[ \hat{U}_{\mathrm{I}}(t),  \left[ \hat{U}_{\mathrm{I}}(t_{1}), \hat{\rho}_{E, \mathrm{I}}(t_{1})\right] \right] =0.
\end{align}

To proceed, we now make some assumptions about impurity scattering. First, we assume that the impurities are uncorrelated and they are of the same type, i.e., $U \left( {\bf r} \right) = \sum_{i} V \left( {\bf r} - {\bf R}_{i} \right)$, where $V \left( {\bf r} \right)$ is the single-impurity potential, and ${\bf R}_{i}$ is the position of the $i$-th impurity~\cite{Kohn-Luttinger1957,Culcer2017}. Second, we assume that the impurities are evenly distributed so that the translational symmetry remains intact at the presence of impurities, and that crystal momentum ${\bf k}$ is still a good quantum number. In practice, this is achieved by averaging over different configurations of the impurities~\cite{Kohn-Luttinger1957,Culcer2017}, see Appendix \ref{sec:disorder-averaging} for details. The spatially homogeneous semiclassical Boltzmann equation in a uniform electric field with symmetric scattering can be obtained by taking the impurity average over Eq.~(\ref{eq:Liouville-3}). Here, the crystal momentum has to be conserved. For example, the third term on the left-hand side of Eq.~(\ref{eq:Liouville-3}) contains a single $\hat{U}$, and the conservation of crystal momentum dictates that it alone cannot change the momentum. Thus, this trivial term does not contribute to scattering, and it is absorbed into the band dispersion~\cite{Kohn-Luttinger1957,Culcer2017}; see Appendix \ref{sec:disorder-averaging} for details. After averaging, we have
\begin{align}
\label{eq:Liouville-4}
\frac{d}{dt} \hat{\rho}_{E, \mathrm{I}}(t) + \frac{i}{\hbar} 
\left[ \hat{H}_{E, I} (t), \hat{\rho}_{0} \right] + \mathcal{J} \left( t \right) = 0, \\
\label{eq:J-t}
\mathcal{J} \left( t \right) = \frac{1}{\hbar^{2}} \int_{-\infty}^{t} d t_{1} \left[ \hat{U}_{\mathrm{I}}(t),  \left[ \hat{U}_{\mathrm{I}}(t_{1}), \hat{\rho}_{E, \mathrm{I}}(t_{1})\right] \right] . 
\end{align}
Equation~(\ref{eq:Liouville-4}) will leads us to the Boltzmann equation. The first term in Eq.~(\ref{eq:Liouville-4}) is linked to the time derivative of the density matrix,
\begin{align}
\label{eq:d-dt}
&\langle m, {\bf k}, t \vert \frac{d}{dt} \hat{\rho}_{E, \mathrm{I}}(t) \vert m^{\prime}, {\bf k}, t \rangle  \nonumber \\
= & \frac{d}{dt} \rho_{ {\bf k} }^{ m m^{ \prime } } \left( t \right) + \frac{i}{\hbar} \left( \varepsilon_{ {\bf k} }^{ m } - \varepsilon_{ {\bf k} }^{ m^{\prime} } \right) \rho_{ {\bf k} }^{ m m^{ \prime } } \left( t \right)  ,
\end{align}
where $\vert m, {\bf k}, t \rangle \equiv e^{ \frac{i}{\hbar} \hat{H}_{0} t} \vert m, {\bf k}  \rangle$, and $\rho_{ {\bf k} }^{ m m^{ \prime } } \left( t \right) \equiv \langle m, {\bf k}, t \vert \hat{\rho}_{E, \mathrm{I}} \left( t \right) \vert m^{\prime}, {\bf k}, t \rangle $. As is stated above, the system is translationally invariant, and we consider only matrix elements diagonal in ${\bf k}$ in the following. Regarding the band number, we separate the density matrix into a diagonal part $f^{m}_{ {\bf k} } \left( t \right)$ and an interband (off-diagonal) part $S^{m m^{\prime} }_{ {\bf k} } \left( t \right)$, i.e., $\rho_{ {\bf k} }^{ m m^{ \prime } } \left( t \right) = \delta_{ m m^{\prime} } f^{m}_{ {\bf k} } \left( t \right) + \left( 1 - \delta_{ m m^{\prime} } \right) S^{m m^{\prime} }_{ {\bf k} } \left( t \right)$. The diagonal part corresponds to the semiclassical distribution function, $f^{m}_{ {\bf k} } \left( t \right) \equiv \rho_{ {\bf k} }^{ m m } \left( t \right)$, which satisfies $\frac{d}{dt} f^{m}_{ {\bf k} } \left( t \right) =  \langle m, {\bf k}, t \vert \frac{d}{dt} \hat{\rho}_{E, \mathrm{I}}(t) \vert m, {\bf k}, t \rangle$. From the perspective of  the density matrix, the diagonal part $f^{m}_{ {\bf k} } \left( t \right)$ does not have the full information of the state of the system, and the gauge-dependent interband part $S^{m m^{\prime} }_{ {\bf k} } \left( t \right)$, which embodies interband coherence and is related to quantum geometry, is essential. Semiclassically, the state of the system is described by $f^{m}_{ {\bf k} } \left( t \right)$, while the information in $S^{m m^{\prime} }_{ {\bf k} } \left( t \right)$ is reflected in the anomalous velocity induced by the Berry curvature and the side-jump corrections. 

The second term in Eq.~(\ref{eq:Liouville-4}) is the driving term, It can be written with the position operator ${\bf r}$ as~\cite{Kohn-Luttinger1957,Blount1962,Culcer2017} (see Appendix \ref{sec:r-operator} for details) 
\begin{align}
\label{eq:driving}
& \langle m, {\bf k}, t \vert \frac{i}{\hbar} \left[ \hat{H}_{E, I} (t), \hat{\rho}_{0} \right] \vert m^{\prime}, {\bf k}, t \rangle \nonumber \\
= & - \frac{e}{\hbar} \delta_{ m m^{\prime} } {\boldsymbol{\mathcal{E}}}(t) \cdot \nabla_{ {\bf k} } f_{0, {\bf k} }^{m } + i \frac{e}{\hbar} {\boldsymbol{\mathcal{E}}}(t) \cdot {\bf A}_{ {\bf k} }^{m m^{\prime} } \left( f_{0, {\bf k} }^{m^{\prime} } - f_{0, {\bf k} }^{m} \right),
\end{align}
where ${\bf A}_{ {\bf k} }^{m m^{\prime} } = i \int d{\bf r} u^{*}_{m, {\bf k} } \left( {\bf r} \right) \nabla_{ {\bf k} } u_{ m^{\prime}, {\bf k} } \left( {\bf r} \right)$ is the Berry connection, with $u_{ m, {\bf k} }$ being the periodic part of Bloch wave function. The first term on the right-hand side of Eq.~(\ref{eq:driving}) is diagonal in the band indices, which is the same as the Fermi surface shift induced by an applied electric field in the semiclassical picture. The second term is nonzero only for $m \ne m^{\prime}$, and it embodies the interband coherence and quantum geometry.

\begin{table}
\begin{tabular}{l c }
\hline\hline
Term & Orders in $n_{i}$ and $V$  \\
\hline  \\[-2.8ex]
$f_{ {\bf k} }^{m} \left( t \right)$ & $ \left( n_{i} V^{2} \right)^{ -1 } $ \\
$S_{ {\bf k} }^{ m m^{ \prime } } \left( t \right)$ & $ \left( n_{i} V^{2} \right)^{ 0 } $ \\
$J_{ f, {\bf k} }^{ m m } \left( t \right)$ and $J_{ f, {\bf k} }^{ m m^{ \prime } } \left( t \right)$ & $\left( n_{i} V^{2} \right)^{ 0 }$\\
$J_{ S, {\bf k} }^{ m m } \left( t \right)$ and $J_{ S, {\bf k} }^{ m m^{ \prime } } \left( t \right)$ & $\left( n_{i} V^{2} \right)^{ 1 }$\\[2.5pt] 
\hline \hline
\end{tabular}
\caption{Orders of the terms in Eqs.~(\ref{eq:diagonal-1}) and (\ref{eq:off-diagonal-1}) in the impurity density $n_{i}$ and the scattering potential $V$. Here, we consider only the leading-order contribution of the terms. }
\label{Table:Order}
\end{table}

The last term in Eq.~(\ref{eq:Liouville-4}), $\mathcal{J} \left( t \right)$, will give us the collision integral. In the interaction picture, its matrix element, ${J}_{ {\bf k} }^{ m m^{\prime} } \left( t \right) \equiv \langle m, {\bf k}, t \vert \mathcal{J} \left( t \right) \vert m^{\prime}, {\bf k}, t \rangle$, reads
\begin{align}
\label{eq:scattering}
& {J}_{ {\bf k} }^{ m m^{\prime} } \left( t \right) \nonumber \\
= & - \frac{ i }{ \hbar } \sum_{ m^{ \prime \prime}, m^{ \prime \prime \prime } } \int d{\bf k}^{ \prime } \nonumber \\
& \bigg\{ \frac{ U_{ {\bf k} {\bf k}^{ \prime } }^{ m m^{ \prime \prime} } U_{ {\bf k}^{ \prime } {\bf k} }^{ m^{ \prime \prime} m^{ \prime \prime \prime } } \rho_{ {\bf k} }^{ m^{ \prime \prime \prime } m^{ \prime } } \left( t \right) - U_{ {\bf k} {\bf k}^{ \prime } }^{ m m^{ \prime \prime} } \rho_{ {\bf k}^{ \prime } }^{ m^{ \prime \prime} m^{ \prime \prime \prime } } \left( t \right)  U_{ {\bf k}^{ \prime} {\bf k} }^{ m^{ \prime \prime \prime } m^{ \prime } } }{ \varepsilon_{ {\bf k}^{ \prime} }^{ m^{ \prime \prime} } - \varepsilon_{ {\bf k} }^{ m^{ \prime } } - i \hbar \eta } \nonumber \\
& + \frac{ \rho_{ {\bf k} }^{ m m^{ \prime \prime \prime} }  \left( t \right) U_{ {\bf k} {\bf k}^{ \prime } }^{ m^{ \prime \prime \prime} m^{ \prime \prime } } U_{ {\bf k}^{ \prime } {\bf k} }^{ m^{ \prime \prime } m^{ \prime } } - U_{ {\bf k} {\bf k}^{ \prime} }^{ m m^{ \prime \prime \prime } } \rho_{ {\bf k}^{ \prime} }^{ m^{ \prime \prime \prime} m^{ \prime \prime } } \left( t \right)  U_{ {\bf k}^{ \prime} {\bf k} }^{ m^{ \prime \prime } m^{ \prime } } }{  \varepsilon_{ {\bf k} }^{ m }- \varepsilon_{ {\bf k}^{ \prime } }^{ m^{ \prime \prime } } - i \hbar \eta } \bigg\},
\end{align}
where $U_{ {\bf k} {\bf k}^{ \prime } }^{ m m^{ \prime } } \equiv \langle m, {\bf k}, t \vert \hat{U}_{\mathrm{I}}(t) \vert m^{\prime}, {\bf k}^{ \prime }, t \rangle$. We have already utilized the assumption that $\hbar \omega \ll \vert \varepsilon_{ {\bf k} }^{ m } - \varepsilon_{ {\bf k} }^{ m^{\prime} } \vert$ $\left(m \ne m^{\prime}\right)$. 

As stated earlier, the impurity potential $U$ in the integrand in Eq.~(\ref{eq:scattering}) will be averaged over all possible impurity configurations. In principle, this is very hard, as $\rho_{ {\bf k} }^{ m  m^{ \prime } } \left( t \right)$ inside the integrand can also contain the impurity potential $U$. Here, we take the first Born approximation, and the two $U$'s in each term in the curly brackets in Eq.~(\ref{eq:scattering}) are paired for the ensemble average, $\langle U_{ {\bf k} {\bf k}^{ \prime } }^{ m m^{ \prime} } U_{ {\bf k}^{ \prime} {\bf k} }^{ m^{ \prime \prime } m^{ \prime \prime \prime} } \rangle = n_{i} V_{ {\bf k} {\bf k}^{ \prime } }^{ m m^{ \prime} } V_{ {\bf k}^{ \prime} {\bf k} }^{ m^{ \prime \prime } m^{ \prime \prime \prime} }$, where the angle brackets $\langle \dots \rangle$ indicate the average over impurity configurations, $n_{i}$ is the density of the impurities, $V_{ {\bf k} {\bf k}^{ \prime } }^{ m m^{ \prime} } \equiv \langle m, {\bf k} \vert \hat{V} \vert m^{\prime}, {\bf k}^{ \prime } \rangle$, and $\hat{V}$ is the single-impurity potential; see Appendix \ref{sec:disorder-averaging} for details. Then, we can rewrite Eq.~(\ref{eq:scattering}) as
\begin{align}
\label{eq:scattering-2}
& {J}_{ {\bf k} }^{ m m^{\prime} } \left( t \right) \nonumber \\
= & - \frac{ i }{ \hbar } n_{i} \sum_{ m^{ \prime \prime}, m^{ \prime \prime \prime } } \int d{\bf k}^{ \prime } \nonumber \\
& \bigg\{ \frac{ V_{ {\bf k} {\bf k}^{ \prime } }^{ m m^{ \prime \prime} } V_{ {\bf k}^{ \prime } {\bf k} }^{ m^{ \prime \prime} m^{ \prime \prime \prime } } \rho_{ {\bf k} }^{ m^{ \prime \prime \prime } m^{ \prime } } \left( t \right) - V_{ {\bf k} {\bf k}^{ \prime } }^{ m m^{ \prime \prime} } \rho_{ {\bf k}^{ \prime } }^{ m^{ \prime \prime} m^{ \prime \prime \prime } } \left( t \right)  V_{ {\bf k}^{ \prime} {\bf k} }^{ m^{ \prime \prime \prime } m^{ \prime } } }{ \varepsilon_{ {\bf k}^{ \prime} }^{ m^{ \prime \prime} } - \varepsilon_{ {\bf k} }^{ m^{ \prime } } - i \hbar \eta } \nonumber \\
& + \frac{ \rho_{ {\bf k} }^{ m m^{ \prime \prime \prime} }  \left( t \right) V_{ {\bf k} {\bf k}^{ \prime } }^{ m^{ \prime \prime \prime} m^{ \prime \prime } } V_{ {\bf k}^{ \prime } {\bf k} }^{ m^{ \prime \prime } m^{ \prime } } - V_{ {\bf k} {\bf k}^{ \prime} }^{ m m^{ \prime \prime \prime } } \rho_{ {\bf k}^{ \prime} }^{ m^{ \prime \prime \prime} m^{ \prime \prime } } \left( t \right) V_{ {\bf k}^{ \prime} {\bf k} }^{ m^{ \prime \prime } m^{ \prime } } }{  \varepsilon_{ {\bf k} }^{ m }- \varepsilon_{ {\bf k}^{ \prime } }^{ m^{ \prime \prime } } - i \hbar \eta } \bigg\}.
\end{align}
Roughly speaking, with the first Born approximation, impurity scattering introduces a factor of $n_{i} V^{2}$ here.

We will see in the following that the diagonal component $f^{m}_{ {\bf k} } \left( t \right)$ and the off-diagonal component $S^{m m^{\prime} }_{ {\bf k} } \left( t \right)$ have different dependences on the impurity density $n_{i}$ and the scattering potential $V$. To facilitate evaluation, we break $J_{ {\bf k} }^{ m m^{\prime} } \left( t \right)$ into two parts, $J_{ {\bf k} }^{ m m^{\prime} } \left( t \right) = J_{ f, {\bf k} }^{ m m^{\prime} } \left( t \right) + J_{ S, {\bf k} }^{ m m^{\prime} } \left( t \right)$, with $J_{ f, {\bf k} }^{ m m^{\prime} } \left( t \right)$ and $J_{ S, {\bf k} }^{ m m^{\prime} } \left( t \right)$ containing the contributions of $f^{m}_{ {\bf k} } \left( t \right)$ and $S^{m m^{\prime} }_{ {\bf k} } \left( t \right)$, respectively:
\begin{align}
\label{eq:J-f}
J_{ f, {\bf k} }^{ m m^{\prime} } \left( t \right) = & - \frac{ i }{ \hbar } n_{i} \sum_{ m^{ \prime \prime} } \int d{\bf k}^{ \prime} V_{ {\bf k} {\bf k}^{ \prime } }^{ m m^{ \prime \prime } } V_{ {\bf k}^{ \prime} {\bf k} }^{ m^{ \prime \prime } m^{ \prime } }  \nonumber \\
& \times \bigg[ \frac{ f_{ {\bf k} }^{ m^{ \prime } } \left( t \right) - f_{ {\bf k}^{ \prime} }^{ m^{ \prime \prime } } \left( t \right) }{ \varepsilon_{ {\bf k}^{ \prime} }^{ m^{ \prime \prime} } - \varepsilon_{ {\bf k} }^{ m^{ \prime } } - i \hbar \eta } + \frac{ f_{ {\bf k} }^{ m } \left( t \right) - f_{ {\bf k}^{ \prime } }^{ m^{ \prime \prime } } \left( t \right) }{  \varepsilon_{ {\bf k} }^{ m }- \varepsilon_{ {\bf k}^{ \prime } }^{ m^{ \prime \prime } } - i \hbar \eta } \bigg], \\
\label{eq:J-S}
J_{ S, {\bf k} }^{ m m^{\prime} } \left( t \right) = & - \frac{ i }{ \hbar } n_{i} \int d{\bf k}^{ \prime}  \nonumber \\
& \times \bigg[  \sum_{ m^{ \prime \prime} } \sum_{ m^{ \prime \prime \prime } \ne m^{ \prime } } \frac{ V_{ {\bf k} {\bf k}^{ \prime } }^{ m m^{ \prime \prime} } V_{ {\bf k}^{ \prime } {\bf k} }^{ m^{ \prime \prime} m^{ \prime \prime \prime } } S_{ {\bf k} }^{ m^{ \prime \prime \prime } m^{ \prime } } \left( t \right)}{ \varepsilon_{ {\bf k}^{ \prime} }^{ m^{ \prime \prime} } - \varepsilon_{ {\bf k} }^{ m^{ \prime } } - i \hbar \eta } \nonumber \\
& - \sum_{ m^{ \prime \prime} \ne m^{ \prime \prime \prime } } \frac{ V_{ {\bf k} {\bf k}^{ \prime } }^{ m m^{ \prime \prime} } S_{ {\bf k}^{ \prime } }^{ m^{ \prime \prime} m^{ \prime \prime \prime } } \left( t \right)  V_{ {\bf k}^{ \prime} {\bf k} }^{ m^{ \prime \prime \prime } m^{ \prime } } }{ \varepsilon_{ {\bf k}^{ \prime} }^{ m^{ \prime \prime} } - \varepsilon_{ {\bf k} }^{ m^{ \prime } } - i \hbar \eta } \nonumber \\
& + \sum_{ m^{ \prime \prime} } \sum_{ m^{ \prime \prime \prime } \ne m } \frac{ S_{ {\bf k} }^{ m m^{ \prime \prime \prime} }  \left( t \right) V_{ {\bf k} {\bf k}^{ \prime } }^{ m^{ \prime \prime \prime} m^{ \prime \prime } } V_{ {\bf k}^{ \prime } {\bf k} }^{ m^{ \prime \prime } m^{ \prime } }}{  \varepsilon_{ {\bf k} }^{ m }- \varepsilon_{ {\bf k}^{ \prime } }^{ m^{ \prime \prime } } - i \hbar \eta }  \nonumber \\
& - \sum_{ m^{ \prime \prime} \ne m^{ \prime \prime \prime } }\frac{ V_{ {\bf k} {\bf k}^{ \prime} }^{ m m^{ \prime \prime \prime } } S_{ {\bf k}^{ \prime} }^{ m^{ \prime \prime \prime} m^{ \prime \prime } } \left( t \right)  V_{ {\bf k}^{ \prime} {\bf k} }^{ m^{ \prime \prime } m^{ \prime } } }{  \varepsilon_{ {\bf k} }^{ m }- \varepsilon_{ {\bf k}^{ \prime } }^{ m^{ \prime \prime } } - i \hbar \eta } \bigg].
\end{align}

Now, we write separate equations for the diagonal part and the off-diagonal part of Eq.~(\ref{eq:Liouville-4}),  
\begin{align}
\label{eq:diagonal-1}
& \frac{d}{dt} f_{ {\bf k} }^{m} \left( t \right) - \frac{e}{\hbar} {\boldsymbol{\mathcal{E}}}(t) \cdot \nabla_{ {\bf k} } f_{0, {\bf k} }^{m } + J_{ f, {\bf k} }^{ m m } \left( t \right) + J_{ S, {\bf k} }^{ m m } \left( t \right) = 0, \\
\label{eq:off-diagonal-1}
& \frac{d}{dt} S_{ {\bf k} }^{ m m^{ \prime } } \left( t \right) + \frac{i}{\hbar} \left( \varepsilon_{ {\bf k} }^{ m } - \varepsilon_{ {\bf k} }^{ m^{\prime} } \right) S_{ {\bf k} }^{ m m^{ \prime } } \left( t \right) \nonumber \\
+ & i \frac{e}{\hbar} {\boldsymbol{\mathcal{E}}}(t) \cdot {\bf A}_{ {\bf k} }^{m m^{\prime} } \left( f_{0, {\bf k} }^{m^{\prime} } - f_{0, {\bf k} }^{m} \right) + J_{ f, {\bf k} }^{ m m^{\prime} } \left( t \right) + J_{ S, {\bf k} }^{ m m^{\prime} } \left( t \right) = 0.
\end{align}
The scattering terms $J_{ f, {\bf k} }^{ m m } \left( t \right)$, $J_{ S, {\bf k} }^{ m m } \left( t \right)$, $J_{ S, {\bf k} }^{ m m^{\prime} } \left( t \right)$ and $J_{ S, {\bf k} }^{ m m^{\prime} } \left( t \right)$ make Eqs.~(\ref{eq:diagonal-1}) and (\ref{eq:off-diagonal-1}) two integral equations, which are difficult to solve. We adopt a systematic approach and solve them in a perturbative fashion~\cite{Kohn-Luttinger1957}. We assume impurity scattering is weak and $n_{i} V^{2}$ is a small number, group the terms in powers of $n_{i} V^{2}$, decompose the equations according to the orders in $n_{i} V^{2}$ and solve them. The orders of the terms are summarized in Table~\ref{Table:Order} and are explained below.

We start from Eq.~(\ref{eq:diagonal-1}). Here, the second term and the fourth term are the driving terms. The second term, $- \frac{e}{\hbar} {\boldsymbol{\mathcal{E}}}(t) \cdot \nabla_{ {\bf k} } f_{0, {\bf k} }^{m }$, describes Fermi surface shift induced by the applied electric field semiclassically, and it is independent of impurity scattering; i.e., it is at order $\left( n_{i} V^{2} \right)^{ 0 }$. The fourth term in Eq.~(\ref{eq:diagonal-1}), $J_{ S, {\bf k} }^{ m m } \left( t \right)$, contains the effect of scattering and interband coherence on the semiclassical distribution function. According to Eq.~(\ref{eq:J-S}), it is at order $\left( n_{i} V^{2} \right) S_{ {\bf k} }^{ m m^{ \prime } } \left( t \right)$, as we take the first-Born approximation and impurity scattering contributes a factor of $n_{i} V^{2}$ after we take ensemble average. The third term, $J_{ f, {\bf k} }^{ m m } \left( t \right)$, gives the familiar collision integral, 
\begin{align} 
\label{eq:J-f-k}
J_{ f, {\bf k} }^{ m m } \left( t \right) = & \int d{\bf k}^{\prime} w^{ \mathrm{S}, m }_{ {\bf k} {\bf k}^{\prime} } \left[ f_{ {\bf k} }^{m} \left( t \right) - f_{ {\bf k}^{\prime} }^{m} \left( t \right) \right],\\
\label{eq:w-S}
w^{ \mathrm{S}, m }_{ {\bf k} {\bf k}^{\prime} } = &\frac{2 \pi}{\hbar} n_{i} V_{ {\bf k} {\bf k}^{ \prime } }^{ m m } V_{ {\bf k}^{ \prime} {\bf k} }^{ m m } \delta \left( \varepsilon_{ {\bf k} }^{ m } - \varepsilon_{ {\bf k}^{\prime} }^{ m } \right) ,
\end{align}
where $w^{ \mathrm{S}, m }_{ {\bf k} {\bf k}^{\prime} }$ is the symmetric scattering rate, which agrees with Fermi's golden rule, and it satisfies $w^{ \mathrm{S}, m }_{ {\bf k} {\bf k}^{\prime} } = w^{ \mathrm{S}, m }_{ {\bf k}^{\prime} {\bf k} }$. Note that following Eq.~(\ref{eq:J-f}), in general we have $J_{ f, {\bf k} }^{ m m } \left( t \right) = \frac{2 \pi}{\hbar} n_{i} \int d{\bf k}^{\prime} V_{ {\bf k} {\bf k}^{ \prime } }^{ m m^{ \prime } } V_{ {\bf k}^{ \prime} {\bf k} }^{ m^{ \prime } m } \left[ f_{ {\bf k} }^{m} \left( t \right) - f_{ {\bf k}^{\prime} }^{ m^{ \prime } } \left( t \right) \right] \delta \left( \varepsilon_{ {\bf k} }^{ m } - \varepsilon_{ {\bf k}^{\prime} }^{ m^{ \prime } } \right)$. According to our assumption that the bands are separated, the $\delta$ function here implies $m^{\prime} = m$, and we have the scattering rate in Eq.~(\ref{eq:w-S}). Following the same logic, $J_{ f, {\bf k} }^{ m m } \left( t \right)$ is at order $\left( n_{i} V^{2} \right) f_{ {\bf k} }^{ m } \left( t \right)$. The first term $\frac{d}{dt} f_{ {\bf k} }^{m} \left( t \right)$ has a magnitude of $\omega f_{ {\bf k} }^{m} \left( t \right)$. In electric transport, typically the frequencies are low, $\omega < \tau^{-1}$, where $\tau$ is the relaxation time and $\tau^{-1} \propto w^{ \mathrm{S}, m }_{ {\bf k} {\bf k}^{\prime} }$~\cite{Ashcroft1976}, i.e., $\tau \propto \left( n_{i} V^{2} \right)^{-1}$. In the DC limit, $\omega \tau \ll 1$ and $\frac{d}{dt} f_{ {\bf k} }^{m} \left( t \right)$ is negligible compared to $J_{ f, {\bf k} }^{ m m } \left( t \right)$. Here we generously allow $\omega \tau \lesssim 1$, taking $J_{ f, {\bf k} }^{ m m } \left( t \right)$ and $\frac{d}{dt} f_{ {\bf k} }^{m} \left( t \right)$ to be of same order, and we can write $\frac{d}{dt} f_{ {\bf k} }^{m} \left( t \right) + J_{ f, {\bf k} }^{ m m } \left( t \right) = \frac{e}{\hbar} {\boldsymbol{\mathcal{E}}}(t) \cdot \nabla_{ {\bf k} } f_{0, {\bf k} }^{m } - J_{ S, {\bf k} }^{ m m } \left( t \right)$, i.e., $\left( n_{i} V^{2} \right) f_{ {\bf k} }^{ m } \left( t \right) \sim \mathcal{O} \left( \left( n_{i} V^{2} \right)^{ 0 } \right) + \mathcal{O} \left( n_{i} V^{2} S_{ {\bf k} }^{ m m^{ \prime } } \left(  t \right) \right) $. For now, we have $f_{ {\bf k} }^{ m } \left( t \right) \sim \mathcal{O} \left( \left( n_{i} V^{2} \right)^{ -1 } \right) + \mathcal{O} \left( S_{ {\bf k} }^{ m m^{ \prime } } \left( t \right) \right) $, where the two terms are contributed by $\frac{e}{\hbar} {\boldsymbol{\mathcal{E}}}(t) \cdot \nabla_{ {\bf k} } f_{0, {\bf k} }^{m }$ and $J_{ S, {\bf k} }^{ m m } \left( t \right)$, respectively. 

The order of $S_{ {\bf k} }^{ m m^{ \prime } } \left( t \right)$ is analyzed with Eq.~(\ref{eq:off-diagonal-1}). The first term, $\frac{d}{dt} S_{ {\bf k} }^{ m m^{ \prime } } \left( t \right)$, is negligible compared to $\frac{i}{\hbar} \left( \varepsilon_{ {\bf k} }^{ m } - \varepsilon_{ {\bf k} }^{ m^{\prime} } \right) S_{ {\bf k} }^{ m m^{ \prime } } \left( t \right)$, as we take the low-frequency limit $\hbar \omega \ll \vert \varepsilon_{ {\bf k} }^{ m } - \varepsilon_{ {\bf k} }^{ m^{\prime} } \vert$ $\left(m \ne m^{\prime}\right)$ in the semiclassical regime. Therefore, we have the formal solution of Eq.~(\ref{eq:off-diagonal-1}): $S_{ {\bf k} }^{ m m^{ \prime } } \left( t \right) = e {\boldsymbol{\mathcal{E}}} (t) \cdot {\bf A}_{ {\bf k} }^{m m^{\prime} } \frac{ f_{0, {\bf k} }^{m} - f_{0, {\bf k} }^{m^{\prime} } }{ \varepsilon_{ {\bf k} }^{ m } - \varepsilon_{ {\bf k} }^{ m^{\prime} } } + i \hbar \frac{ 1 }{ \varepsilon_{ {\bf k} }^{ m } - \varepsilon_{ {\bf k} }^{ m^{\prime} } } J_{ f, {\bf k} }^{ m m^{\prime} } \left( t \right) + i \hbar \frac{ 1 }{ \varepsilon_{ {\bf k} }^{ m } - \varepsilon_{ {\bf k} }^{ m^{\prime} } } J_{ S, {\bf k} }^{ m m^{\prime} } \left( t \right)$. The first part, $e {\boldsymbol{\mathcal{E}}} (t) \cdot {\bf A}_{ {\bf k} }^{m m^{\prime} } \frac{ f_{0, {\bf k} }^{m} - f_{0, {\bf k} }^{m^{\prime} } }{ \varepsilon_{ {\bf k} }^{ m } - \varepsilon_{ {\bf k} }^{ m^{\prime} } }$, embodies field induced interband coherence. It is free of scattering, and it is at order $\left( n_{i} V^{2} \right)^{ 0 }$. The second part $i \hbar \frac{ 1 }{ \varepsilon_{ {\bf k} }^{ m } - \varepsilon_{ {\bf k} }^{ m^{\prime} } } J_{ f, {\bf k} }^{ m m^{\prime} } \left( t \right)$ involves interband coherence originating from the scattering of the distribution function. The term is at order $\left( n_{i} V^{2} \right) f_{ {\bf k} }^{ m } \left( t \right)$ in the first Born approximation [see Eq.~(\ref{eq:J-f})]. Since we have $f_{ {\bf k} }^{ m } \left( t \right) \sim \mathcal{O} \left( \left( n_{i} V^{2} \right)^{ -1 } \right) + \mathcal{O} \left( S_{ {\bf k} }^{ m m^{ \prime } } \left( t \right) \right) $, $i \hbar \frac{ 1 }{ \varepsilon_{ {\bf k} }^{ m } - \varepsilon_{ {\bf k} }^{ m^{\prime} } } J_{ f, {\bf k} }^{ m m^{\prime} } \sim \mathcal{O} \left( \left( n_{i} V^{2} \right)^{ 0 } \right) + \mathcal{O} \left(  n_{i} V^{2} S_{ {\bf k} }^{ m m^{ \prime } } \left( t \right) \right)$. The last part of the formal solution, $i \hbar \frac{ 1 }{ \varepsilon_{ {\bf k} }^{ m } - \varepsilon_{ {\bf k} }^{ m^{\prime} } } J_{ S, {\bf k} }^{ m m^{\prime} } \left( t \right)$, is also related to scattering. In the same vein, it is at order $\left( n_{i} V^{2} \right) S_{ {\bf k} }^{ m m^{ \prime } } \left( t \right)$, following the expression in Eq.~(\ref{eq:J-S}). As a result, we have $S_{ {\bf k} }^{ m m^{ \prime } } \left( t \right) \sim \mathcal{O} \left( \left( n_{i} V^{2} \right)^{ 0 } \right) + \mathcal{O} \left(  n_{i} V^{2} S_{ {\bf k} }^{ m m^{ \prime } } \left( t \right) \right)$. The leading-order contribution to $S_{ {\bf k} }^{ m m^{ \prime } } \left( t \right)$ is at order $\left( n_{i} V^{2} \right)^{ 0 }$, and the leading-order part of $f_{ {\bf k} }^{ m } \left( t \right)$ is at order $\left( n_{i} V^{2} \right)^{ -1 }$. The orders of the other terms are determined accordingly, see Table~\ref{Table:Order}.  

To investigate electric transport, we solve the diagonal part $f_{ {\bf k} }^{ m } \left( t \right)$ and the off-diagonal component $S_{ {\bf k} }^{ m m^{ \prime } } \left( t \right)$ to the same order, at order $\left( n_{i} V^{2} \right)^{ 0 }$. To this end, we reorganize Eq.~(\ref{eq:diagonal-1}) according to orders of $n_{i} V^{2}$. We write the solution to Eq.~(\ref{eq:diagonal-1}) as
\begin{equation}
\label{eq:f_k-f_1-f_sj}
f_{ {\bf k} }^{m} \left( t \right) = f_{ 1, {\bf k} }^{m} \left( t \right) + f_{ \mathrm{sj}, {\bf k} }^{m} \left( t \right),
\end{equation} 
where $f_{ 1, {\bf k} }^{m} \left( t \right)$ is the leading-order term, at order $\left( n_{i} V^{2} \right)^{ -1 }$. The other term $f_{ \mathrm{sj}, {\bf k} }^{m} \left( t \right)$ is at order $\left( n_{i} V^{2} \right)^{ 0 }$, and it is akin to the semiclassical side-jump distribution function, as we will see in Sec.~\ref{sec:side-jump}. With Eq.~(\ref{eq:f_k-f_1-f_sj}), we can rewrite $J_{ f, {\bf k} }^{ m m } \left( t \right)$ in Eq.~(\ref{eq:J-f-k}) as $J_{ f, {\bf k} }^{ m m } \left( t \right) = I_{ 1, {\bf k} }^{ m } \left( t \right) + I_{ \mathrm{sj}, {\bf k} }^{ m } \left( t \right)$, where 
\begin{align}
\label{eq:I-1}
I_{ 1, {\bf k} }^{ m } \left( t \right) &= \int d{\bf k}^{\prime} w^{ \mathrm{S}, m }_{ {\bf k} {\bf k}^{\prime} } \left[ f_{ 1, {\bf k} }^{m} \left( t \right) - f_{ 1, {\bf k}^{\prime} }^{m} \left( t \right) \right], \\
\label{eq:I-sj}
I_{ \mathrm{sj}, {\bf k} }^{ m } \left( t \right) &= \int d{\bf k}^{\prime} w^{ \mathrm{S}, m }_{ {\bf k} {\bf k}^{\prime} } \left[ f_{ \mathrm{sj}, {\bf k} }^{m} \left( t \right) - f_{ \mathrm{sj}, {\bf k}^{\prime} }^{m} \left( t \right) \right].
\end{align}
It is easy to see that $I_{ 1, {\bf k} }^{ m } \left( t \right)$ and $I_{ \mathrm{sj}, {\bf k} }^{ m } \left( t \right)$ are at order $\left( n_{i} V^{2} \right)^{ 0 }$ and $\left( n_{i} V^{2} \right)^{ 1 }$, respectively. Now, we can separate Eq.~(\ref{eq:diagonal-1}) into a leading-order equation and a next-to-leading order one. The leading-order equation reads 
\begin{equation}
\label{eq:diagonal-2}
\frac{d}{dt} f_{ 1, {\bf k} }^{m} \left( t \right) - \frac{e}{\hbar} {\boldsymbol{\mathcal{E}}}(t) \cdot \nabla_{ {\bf k} } f_{0, {\bf k} }^{m } + I_{ 1, {\bf k} }^{ m } \left( t \right) = 0.
\end{equation}
Equation~(\ref{eq:diagonal-2}) is exactly the same as the semiclassical Boltzmann equation in a homogenous system in an applied uniform electric field. Although this equation is frequently set up intuitively in textbooks, we here show that it can be derived directly from the density matrix. The next-to-leading order equation containing $f_{ \mathrm{sj}, {\bf k} }^{m} \left( t \right)$ is given in Sec.~\ref{sec:side-jump}, and it leads to the side-jump correction.

We handle Eq.~(\ref{eq:off-diagonal-1}) in the same way, keeping terms to the order of $\left( n_{i} V^{2} \right)^{ 0 }$. As discussed above, $\frac{d}{dt} S_{ 1, {\bf k} }^{ m m^{ \prime } } \left( t \right)$ can be neglected, and $J_{ S, {\bf k} }^{ m m^{\prime} } \left( t \right)$ is a higher-order term (see Table~\ref{Table:Order}). We see from Eqs.~(\ref{eq:J-f}) and (\ref{eq:f_k-f_1-f_sj}) that $J_{ f, {\bf k} }^{ m m^{\prime} } \left( t \right)$ consists of contributions from $f_{ 1, {\bf k} }^{m} \left( t \right)$ and $f_{ \mathrm{sj}, {\bf k} }^{m} \left( t \right)$. Only the part engendered by $f_{ 1, {\bf k} }^{m} \left( t \right)$ is at order $\left( n_{i} V^{2} \right)^{ 0 }$, as $f_{ 1, {\bf k} }^{m} \left( t \right)$ is at order $\left( n_{i} V^{2} \right)^{ -1 }$, and scattering in the first Born approximation generates a factor of $n_{i} V^{2}$. We denote this part as $J_{ f_{1}, {\bf k} }^{ m m^{\prime} } \left( t \right)$,  
\begin{align}
\label{eq:J-f1}
J_{ f_{1}, {\bf k} }^{ m m^{\prime} } \left( t \right) = & - \frac{ i }{ \hbar } n_{i} \sum_{ m^{ \prime \prime} } \int d{\bf k}^{ \prime} V_{ {\bf k} {\bf k}^{ \prime } }^{ m m^{ \prime \prime } } V_{ {\bf k}^{ \prime} {\bf k} }^{ m^{ \prime \prime } m^{ \prime } }  \nonumber \\
& \bigg[ \frac{ f_{ 1, {\bf k} }^{ m^{ \prime } } \left( t \right) - f_{ 1, {\bf k}^{ \prime} }^{ m^{ \prime \prime } } \left( t \right) }{ \varepsilon_{ {\bf k}^{ \prime} }^{ m^{ \prime \prime} } - \varepsilon_{ {\bf k} }^{ m^{ \prime } } - i \hbar \eta } + \frac{ f_{ 1, {\bf k} }^{ m } \left( t \right) - f_{ 1, {\bf k}^{ \prime } }^{ m^{ \prime \prime } } \left( t \right) }{  \varepsilon_{ {\bf k} }^{ m }- \varepsilon_{ {\bf k}^{ \prime } }^{ m^{ \prime \prime } } - i \hbar \eta } \bigg], 
\end{align}

Retaining the terms at order $\left( n_{i} V^{2} \right)^{ 0 }$, we have the leading-order version of Eq.~(\ref{eq:off-diagonal-1}),
\begin{align}
\label{eq:off-diagonal-1-1}
& \frac{i}{\hbar} \left( \varepsilon_{ {\bf k} }^{ m } - \varepsilon_{ {\bf k} }^{ m^{\prime} } \right) S_{ {\bf k} }^{ m m^{ \prime } } \left( t \right) + i \frac{e}{\hbar} {\boldsymbol{\mathcal{E}}}(t) \cdot {\bf A}_{ {\bf k} }^{m m^{\prime} } \left( f_{0, {\bf k} }^{m^{\prime} } - f_{0, {\bf k} }^{m} \right) \nonumber \\
&  + J_{ f_{1}, {\bf k} }^{ m m^{\prime} } \left( t \right) = 0,
\end{align}
with the solution  
\begin{align}
S_{ {\bf k} }^{ m m^{ \prime } } \left( t \right) = & S_{ 1, {\bf k} }^{ m m^{ \prime } } \left( t \right) + S_{ 2, {\bf k} }^{ m m^{ \prime } } \left( t \right), \\
\label{eq:off-diagonal-2}
S_{ 1, {\bf k} }^{ m m^{ \prime } } \left( t \right) = & e {\boldsymbol{\mathcal{E}}} (t) \cdot {\bf A}_{ {\bf k} }^{m m^{\prime} } \frac{ f_{0, {\bf k} }^{m} - f_{0, {\bf k} }^{m^{\prime} } }{ \varepsilon_{ {\bf k} }^{ m } - \varepsilon_{ {\bf k} }^{ m^{\prime} } }, \\
\label{eq:S-2}
S_{ 2, {\bf k} }^{ m m^{ \prime } } \left( t \right) = & i \hbar \frac{ 1 }{ \varepsilon_{ {\bf k} }^{ m } - \varepsilon_{ {\bf k} }^{ m^{\prime} } } J_{ f_{1}, {\bf k} }^{ m m^{\prime} } \left( t \right).
\end{align}
Both $S_{ 1, {\bf k} }^{ m m^{ \prime } } \left( t \right)$ and $S_{ 2, {\bf k} }^{ m m^{ \prime } } \left( t \right)$ are gauge dependent. $S_{ 2, {\bf k} }^{ m m^{ \prime } } \left( t \right)$ is linked to the semiclassical side jump, and we will discuss it in Sec.~\ref{sec:side-jump}. $S_{ 1, {\bf k} }^{ m m^{ \prime } } \left( t \right)$ contains an interband Berry connection which encodes quantum geometry. In addition, it is a Fermi-sea contribution, as there is no derivative of $f_{0, {\bf k} }^{m}$. We will see later that $S_{ 1, {\bf k} }^{ m m^{ \prime } } \left( t \right)$ generates the anomalous velocity induced by Berry curvature.

Equations~(\ref{eq:diagonal-2}) and (\ref{eq:off-diagonal-2}) recover the semiclassical charge transport in a uniform electric field without skew scattering or side jumps. The charge current density is given by ${\bf j} = - e \mathrm{Tr} \left( \hat{\bf v} \hat{\rho} \right)$, where the velocity operator, $\hat{\bf v} = \frac{d \hat{\bf r} }{ d t } = \frac{ i }{ \hbar } \left[ \hat{H}, \hat{ \bf r } \right]$, satisfies
\begin{align}
\label{eq:velocity}
{\bf v}_{ {\bf k} }^{ m m^{ \prime } } & \equiv \langle m, {\bf k} \vert \hat{\bf v} \vert m^{\prime}, {\bf k} \rangle \nonumber \\
& = \frac{ 1 }{ \hbar } \delta_{ m m^{\prime} } \nabla_{ {\bf k} } \varepsilon_{ {\bf k} }^{ m } + \frac{ i }{ \hbar } {\bf A}_{ {\bf k} }^{m m^{\prime} } \left( \varepsilon_{ {\bf k} }^{ m } - \varepsilon_{ {\bf k} }^{ m^{\prime} } \right) .
\end{align}
The first term is diagonal in band indices, and it involves only dispersion. For simplicity, we introduce the short-hand notation ${\bf v}_{ {\bf k} }^{ m } \equiv \frac{ 1 }{ \hbar } \nabla_{ {\bf k} } \varepsilon_{ {\bf k} }^{ m }$. The second term of ${\bf v}_{ {\bf k} }^{ m m^{ \prime } }$ contains the interband Berry connection and is gauge-dependent. The two terms combine with $f_{ {\bf k} }^{m} \left( t \right)$ and $S_{ {\bf k} }^{ m m^{ \prime } } \left( t \right)$, respectively to generate the charge current ${\bf j}$,
\begin{align}
\label{eq:j-four-components}
{\bf j} \left( t \right) = & - e \int d {\bf k} \bigg[ \sum_{ m } {\bf v}_{ {\bf k} }^{ m } f_{ {\bf k} }^{m} \left( t \right) + \sum_{ m^{\prime} \ne m } {\bf v}_{ {\bf k} }^{ m^{ \prime } m } S_{ {\bf k} }^{ m m^{ \prime } } \left( t \right) \bigg] \nonumber \\
= & {\bf j}_{1} \left( t \right) + {\bf j}_{\mathrm{sj-f}} \left( t \right) + {\bf j}_{\mathrm{a}} \left( t \right) + {\bf j}_{\mathrm{sj-v}} \left( t \right).
\end{align}
${\bf v}_{ {\bf k} }^{ m }$ combines with $f_{ 1, {\bf k} }^{m} \left( t \right)$ to produce the conventional current ${\bf j}_{1} \left( t \right)$, 
\begin{align}
\label{eq:j-1}
{\bf j}_{1} \left( t \right) = - e \sum_{ m } \int d {\bf k} \ {\bf v}_{ {\bf k} }^{ m } f_{ 1, {\bf k} }^{m} \left( t \right) .
\end{align}
${\bf v}_{ {\bf k} }^{ m^{ \prime } m }$ and $S_{ 1, {\bf k} }^{ m m^{ \prime } }$ give rise to the anomalous velocity induced by the Berry curvature and the resultant current ${\bf j}_{\mathrm{a}} \left( t \right)$,
\begin{align}
\label{eq:j-a}
{\bf j}_{\mathrm{a}} \left( t \right) & = - e \sum_{ m^{\prime} \ne m } \int d {\bf k} \ {\bf v}_{ {\bf k} }^{ m^{ \prime } m } S_{ 1, {\bf k} }^{ m m^{ \prime } } \nonumber \\
& = - \frac{ e^{ 2 } }{ \hbar } \sum_{ m } \int d {\bf k} f_{ 0, {\bf k} }^{m} {\boldsymbol{\mathcal{E}}} \left( t \right) \times {\bf \Omega}^{ m }_{ {\bf k} }, 
\end{align}
where ${\bf \Omega}^{ m }_{ {\bf k} } = \nabla \times {\bf A}_{ {\bf k} }^{ m m }$ is the Berry curvature of the $m$ th band. Equation~(\ref{eq:j-a}) agrees with results in earlier studies~\cite{Karplus1954,Kohmoto1985,Nagaosa2010}. Although the Berry curvature and the resultant anomalous velocity are usually attributed to a single band semiclassically, the Berry curvature actually traces its origin to interband effects~\cite{DiXiao2010}, as it comes from the interband Berry connection ${\bf A}_{ {\bf k} }^{m m^{\prime} }$ $\left(m \ne m^{\prime}\right)$ here.

The other two components in Eq.~(\ref{eq:j-four-components}), ${\bf j}_{\mathrm{sj-f}} \left( t \right)$ and ${\bf j}_{\mathrm{sj-v}} \left( t \right)$, contain side-jump contribution and quantum corrections, and they are handled in the following section.  

\section{Side jumps and longitudinal corrections}
\label{sec:side-jump}
The diagonal scattering term $J_{ S, {\bf k} }^{ m m } \left( t \right)$ and the off-diagonal term $S_{ 2, {\bf k} }^{ m m^{ \prime } } \left( t \right)$ are related to the semiclassical side-jump corrections to the distribution function and velocity, respectively. Moreover, they also give corrections to the longitudinal charge current, which are absent in the semiclassical treatment. We start with the combination of ${\bf v}_{ {\bf k} }^{ m^{ \prime } m }$ and $S_{ 2, {\bf k} }^{ m m^{ \prime } } \left( t \right)$,
\begin{equation}
{\bf j}_{ \mathrm{sj-v} } \left( t \right) = - e \sum_{ m \ne m^{ \prime } } \int d{\bf k} \ {\bf v}_{ {\bf k} }^{ m^{ \prime } m } S_{ 2, {\bf k} }^{ m m^{ \prime } } \left( t \right).
\end{equation}
${\bf j}_{ \mathrm{sj-v} } \left( t \right)$ involves quantum geometry and impurity scattering, as the off-diagonal velocity ${\bf v}_{ {\bf k} }^{ m^{ \prime } m }$ introduces quantum geometry through the interband Berry connection, and $J_{ f_{1}, {\bf k} }^{ m m^{\prime} } \left( t \right)$ in $S_{ 2, {\bf k} }^{ m m^{ \prime } } \left( t \right)$ describes scattering processes, see Eqs.~(\ref{eq:J-f1}) and (\ref{eq:S-2}). ${\bf j}_{ \mathrm{sj-v} } \left( t \right)$ can be evaluated directly using Eqs.~(\ref{eq:S-2}) and (\ref{eq:velocity}), 
\begin{align}
\label{eq:j-sj-v-2}
&{\bf j}_{ \mathrm{sj-v} } \left( t \right) \nonumber \\
= & - \frac{ e }{ \hbar } n_{i} \sum_{ m, m^{ \prime \prime } } \int d{\bf k} \int d{\bf k}^{ \prime } f_{ 1, {\bf k} }^{m} \left( t \right)   \nonumber \\
& \bigg\{ i \sum_{ m^{ \prime} } \bigg[ \frac{ {\bf A}_{ {\bf k} }^{ m m^{ \prime } } V_{ {\bf k} {\bf k}^{ \prime } }^{ m^{ \prime } m^{ \prime \prime } } - V_{ {\bf k} {\bf k}^{ \prime }  }^{ m m^{ \prime } } {\bf A}_{ {\bf k}^{ \prime } }^{ m^{ \prime } m^{ \prime \prime } } }{ \varepsilon_{ {\bf k}^{ \prime } }^{ m^{ \prime \prime } } - \varepsilon_{ {\bf k} }^{ m } - i \hbar \eta } V_{ {\bf k}^{ \prime } {\bf k} }^{ m^{ \prime \prime } m } \nonumber \\
& + \frac{ V_{ {\bf k}^{ \prime } {\bf k} }^{ m^{ \prime \prime } m^{ \prime } } {\bf A}_{ {\bf k} }^{ m^{ \prime } m }  - {\bf A}_{ {\bf k}^{ \prime } }^{ m^{ \prime \prime } m^{ \prime } } V_{ {\bf k}^{ \prime } {\bf k}  }^{ m^{ \prime } m }  }{ \varepsilon_{ {\bf k} }^{ m } - \varepsilon_{ {\bf k}^{ \prime } }^{ m^{ \prime \prime } } - i \hbar \eta } V_{ {\bf k} {\bf k}^{ \prime } }^{ m m^{ \prime \prime } } \bigg] \nonumber \\
& + 2 \pi \delta \left( \varepsilon_{ {\bf k} }^{ m } - \varepsilon_{ {\bf k}^{\prime} }^{ m^{ \prime \prime } } \right) V_{ {\bf k} {\bf k}^{ \prime }  }^{ m m^{ \prime \prime } } V_{ {\bf k}^{ \prime } {\bf k} }^{ m^{ \prime \prime } m }  \left( {\bf A}_{ {\bf k} }^{ m m } - {\bf A}_{ {\bf k}^{ \prime } }^{ m^{ \prime \prime } m^{ \prime \prime } } \right) \bigg\}.
\end{align}
Equation~(\ref{eq:j-sj-v-2}) can be further simplified if one notices that ${\bf A}_{ {\bf k} }^{ m m^{ \prime } }$ is the off-diagonal component of the position operator $\hat{ \bf r}$. The single-impurity potential $\hat{V}$ can be written as a function of position. Therefore, $\hat{V}$ and $\hat{ \bf r}$ commute, i.e., $[ \hat{V} , \hat{ \bf r}] = 0$. As a result, 
\begin{align}
\label{eq:commutator-r-U}
& \sum_{m^{\prime\prime}} \left( {\bf A}_{ {\bf k} }^{m m^{\prime\prime} } V_{ {\bf k} {\bf k}^{\prime} }^{ m^{\prime\prime} m^{\prime} } - V_{ {\bf k} {\bf k}^{\prime} }^{ m m^{\prime\prime} } {\bf A}_{ {\bf k}^{\prime} }^{m^{\prime\prime} m^{\prime} } \right) \nonumber \\
= & - i \left( \nabla_{ {\bf k} } + \nabla_{ {\bf k}^{\prime} } \right) V_{ {\bf k} {\bf k}^{\prime} }^{ m m^{\prime} },
\end{align}
which can be obtained with the position operator in the length gauge~\cite{Luttinger1958,Blount1962} (see Appendix \ref{sec:r-operator} for the derivation). With Eq.~(\ref{eq:commutator-r-U}), we have 
\begin{align}
\label{eq:j-sj-v-3}
{\bf j}_{ \mathrm{sj-v} } \left( t \right) = & - e \sum_{ m } \int d{\bf k} \left( {\bf v}_{ {\bf k} }^{  \mathrm{sj}, m } + {\bf v}_{ {\bf k} }^{  \mathrm{L}, m } \right) f_{ 1, {\bf k} }^{m} \left( t \right). 
\end{align}
Namely, the current ${\bf j}_{ \mathrm{sj-v} } \left( t \right)$ comes from the combination of  the conventional distribution induced by an applied electric field $f_{ 1, {\bf k} }^{m} \left( t \right)$, the side-jump velocity ${\bf v}_{ {\bf k} }^{ \mathrm{sj}, m }$, and a velocity correction ${\bf v}_{ {\bf k} }^{  \mathrm{L}, m }$, which we name the longitudinal velocity. ${\bf v}_{ {\bf k} }^{ \mathrm{sj}, m }$ satisfies
\begin{align}
\label{eq:v-sj}
{\bf v}_{ {\bf k} }^{ \mathrm{sj}, m } = & \int d{\bf k}^{ \prime } w^{ \mathrm{S}, m }_{ {\bf k}^{\prime} {\bf k} } \delta {\bf r}_{ {\bf k}^{ \prime } {\bf k} }^{m}.
\end{align}
${\bf v}_{ {\bf k} }^{ \mathrm{sj}, m }$ comprises the symmetric scattering rate $w^{ \mathrm{S}, m }_{ {\bf k}^{\prime} {\bf k} }$ and an additional position shift accompanying the scattering from ${\bf k}$ to ${\bf k}^{ \prime }$, $\delta {\bf r}_{ {\bf k}^{ \prime } {\bf k} }^{m}$, 
\begin{align}
\label{eq:delta-r}
\delta {\bf r}_{ {\bf k}^{ \prime } {\bf k} }^{m} = & {\bf A}_{ {\bf k}^{ \prime } }^{ m m } - {\bf A}_{ {\bf k} }^{ m m } - \left( \nabla_{ {\bf k} } + \nabla_{ {\bf k}^{\prime} } \right) \mathrm{arg} V_{ {\bf k}^{ \prime } {\bf k} }^{ m m } .
\end{align}
$\delta {\bf r}_{ {\bf k}^{ \prime } {\bf k} }^{m}$ is a gauge-invariant quantity, and it is related to a gauge-invariant geometric phase~\cite{Sinitsyn2006}. The expressions for ${\bf v}_{ {\bf k} }^{ \mathrm{sj}, m }$ and $\delta {\bf r}_{ {\bf k}^{ \prime } {\bf k} }^{m}$ coincide with the semiclassical result derived with wave packet analysis~\cite{Sinitsyn2006, Sinitsyn2007}. In comparison, the velocity correction ${\bf v}_{ {\bf k} }^{  \mathrm{L}, m }$ is generally not included in the semiclassical Boltzmann transport formalism. ${\bf v}_{ {\bf k} }^{  \mathrm{L}, m }$ comes from interband virtual transition,
\begin{align}
\label{eq:v-sj-L}
{\bf v}_{ {\bf k} }^{ \mathrm{L}, m } = & \frac{ n_{i} }{ \hbar } \sum_{m^{ \prime }} \int d{\bf k}^{ \prime } \frac{ \mathcal{V. P.} }{ \varepsilon_{ {\bf k} }^{ m } - \varepsilon_{ {\bf k}^{ \prime } }^{ m^{ \prime  } } } \left( \nabla_{ {\bf k} } + \nabla_{ {\bf k}^{\prime} } \right) V_{ {\bf k} {\bf k}^{\prime} }^{ m m^{\prime} } V_{ {\bf k}^{\prime} {\bf k} }^{ m^{\prime} m } .
\end{align}
The electric current induced by ${\bf v}_{ {\bf k} }^{ \mathrm{L}, m }$ survives time-reversal symmetry, and the current is longitudinal at the leading order~\cite{Xiao2017}. Consequently, this current is neglected in studies of the anomalous Hall effect~\cite{Xiao2017}, and we call ${\bf v}_{ {\bf k} }^{ \mathrm{L}, m }$ the longitudinal velocity. We expect ${\bf v}_{ {\bf k} }^{ \mathrm{L}, m }$ to play a role in nonlinear longitudinal transport. 

The other side-jump current, ${\bf j}_{\mathrm{sj-f}} \left( t \right)$, comes from the conventional group velocity ${\bf v}_{ {\bf k} }^{ m }$ and the side-jump correction to the distribution function $f_{ \mathrm{sj}, {\bf k} }^{m} \left( t \right)$,
\begin{equation}
\label{eq:j-sj-f}
{\bf j}_{ \mathrm{sj-f} } \left( t \right) = - e \sum_{ m } \int d{\bf k} \ {\bf v}_{ {\bf k} }^{  m } f_{ \mathrm{sj}, {\bf k} }^{m} \left( t \right). 
\end{equation}
Following the analysis in Sec.~\ref{sec:Boltzmann}, $f_{ \mathrm{sj}, {\bf k} }^{m} \left( t \right)$ can be obtained from the next-to-leading-order terms in Eq.~(\ref{eq:diagonal-1}). The equation for these terms reads
\begin{align}
\label{eq:side-jump-diagonal}
\frac{d}{dt}f_{ \mathrm{sj}, {\bf k} }^{m} \left( t \right) + I_{ \mathrm{sj}, {\bf k} }^{ m } \left( t \right) + J_{ S, {\bf k} }^{ m m } \left( t \right) = 0. 
\end{align}
where $I_{ \mathrm{sj}, {\bf k} }^{ m } \left( t \right)$ and $J_{ S, {\bf k} }^{ m m } \left( t \right)$ are defined in Eqs.~(\ref{eq:I-sj}) and (\ref{eq:J-S}) respectively. Note that although $S_{ 1, {\bf k} }^{ m m } \left( t \right)$ and $S_{ 2, {\bf k} }^{ m m } \left( t \right)$ are of the same order, the contribution of $S_{ 2, {\bf k} }^{ m m } \left( t \right)$ to $J_{ S, {\bf k} }^{ m m } \left( t \right)$ involves an interband virtual transition, and it is part of the Gaussian skew-scattering terms, which are handled afterward. Here, we keep only the contribution of $S_{ 1, {\bf k} }^{ m m } \left( t \right)$ to $J_{ S, {\bf k} }^{ m m } \left( t \right)$ and denote it as $J_{ S_{1}, {\bf k} }^{ m m } \left( t \right)$,
\begin{align}
\label{eq:side-jump-diagonal}
\frac{d}{dt}f_{ \mathrm{sj}, {\bf k} }^{m} \left( t \right) + I_{ \mathrm{sj}, {\bf k} }^{ m } \left( t \right) + J_{ S_{1}, {\bf k} }^{ m m } \left( t \right) = 0,
\end{align}
\begin{align}
\label{eq:J-S-1}
J_{ S_{1}, {\bf k} }^{ m m } \left( t \right) = & - \frac{ i }{ \hbar } n_{i} \sum_{ m^{ \prime } } \int d{\bf k}^{ \prime} \bigg[ \sum_{ m^{ \prime \prime } \ne m } \frac{ V_{ {\bf k} {\bf k}^{ \prime } }^{ m m^{ \prime } } V_{ {\bf k}^{ \prime } {\bf k} }^{ m^{ \prime } m^{ \prime \prime } } S_{ 1, {\bf k} }^{ m^{ \prime \prime } m } \left( t \right) }{ \varepsilon_{ {\bf k}^{ \prime} }^{ m^{ \prime } } - \varepsilon_{ {\bf k} }^{ m } - i \hbar \eta } \nonumber \\
& - \sum_{ m^{ \prime \prime} \ne m^{ \prime } } \frac{ V_{ {\bf k} {\bf k}^{ \prime } }^{ m m^{ \prime } } S_{ 1, {\bf k}^{ \prime } }^{ m^{ \prime} m^{ \prime \prime } } \left( t \right)  V_{ {\bf k}^{ \prime} {\bf k} }^{ m^{ \prime \prime } m } }{ \varepsilon_{ {\bf k}^{ \prime} }^{ m^{ \prime } } - \varepsilon_{ {\bf k} }^{ m } - i \hbar \eta } \nonumber \\
& + \sum_{ m^{ \prime \prime } \ne m } \frac{ S_{ 1, {\bf k} }^{ m m^{ \prime \prime} }  \left( t \right) V_{ {\bf k} {\bf k}^{ \prime } }^{ m^{ \prime \prime} m^{ \prime } } V_{ {\bf k}^{ \prime } {\bf k} }^{ m^{ \prime } m }}{  \varepsilon_{ {\bf k} }^{ m }- \varepsilon_{ {\bf k}^{ \prime } }^{ m^{ \prime } } - i \hbar \eta }  \nonumber \\
& - \sum_{ m^{ \prime \prime} \ne m^{ \prime } }\frac{ V_{ {\bf k} {\bf k}^{ \prime} }^{ m m^{ \prime \prime } } S_{ 1, {\bf k}^{ \prime} }^{ m^{ \prime \prime} m^{ \prime } } \left( t \right)  V_{ {\bf k}^{ \prime} {\bf k} }^{ m^{ \prime } m } }{  \varepsilon_{ {\bf k} }^{ m }- \varepsilon_{ {\bf k}^{ \prime } }^{ m^{ \prime } } - i \hbar \eta } \bigg].
\end{align} 
If we substitute Eq.~(\ref{eq:off-diagonal-2}) into Eq.~(\ref{eq:J-S-1}), 
\begin{align}
\label{eq:J-S-2}
J_{ \mathrm{S}_{1}, {\bf k} }^{ m m } \left( t \right) = & - \frac{ i }{ \hbar } n_{i} e {\boldsymbol{\mathcal{E}}} (t) \cdot \sum_{ m^{ \prime } } \int d{\bf k}^{ \prime}   \nonumber \\
& \times \bigg[ \sum_{ m^{ \prime \prime } \ne m } \frac{ V_{ {\bf k} {\bf k}^{ \prime } }^{ m m^{ \prime } } V_{ {\bf k}^{ \prime } {\bf k} }^{ m^{ \prime } m^{ \prime \prime } } {\bf A}_{ {\bf k} }^{ m^{ \prime \prime } m } }{ \varepsilon_{ {\bf k}^{ \prime} }^{ m^{ \prime } } - \varepsilon_{ {\bf k} }^{ m } - i \hbar \eta } \frac{ f_{0, {\bf k} }^{ m^{ \prime \prime } } - f_{0, {\bf k} }^{m } }{ \varepsilon_{ {\bf k} }^{ m^{ \prime \prime } }  - \varepsilon_{ {\bf k} }^{ m } } \nonumber \\
& - \sum_{ m^{ \prime \prime } \ne m^{ \prime } } \frac{ V_{ {\bf k} {\bf k}^{ \prime } }^{ m m^{ \prime } } V_{ {\bf k}^{ \prime} {\bf k} }^{ m^{ \prime \prime } m } {\bf A}_{ {\bf k}^{ \prime} }^{ m^{ \prime } m^{ \prime \prime } } }{ \varepsilon_{ {\bf k}^{ \prime} }^{ m^{ \prime } } - \varepsilon_{ {\bf k} }^{ m } - i \hbar \eta } \frac{ f_{0, {\bf k}^{ \prime } }^{ m^{ \prime } } - f_{0, {\bf k}^{ \prime } }^{ m^{ \prime \prime } } }{ \varepsilon_{ {\bf k}^{ \prime } }^{ m^{ \prime } } - \varepsilon_{ {\bf k}^{ \prime } }^{ m^{ \prime \prime } } } - \mathrm{c. c.} \bigg]. 
\end{align} 
The scattering term $J_{ S, {\bf k} }^{ m m } \left( t \right)$ is a mixture of impurity scattering, quantum geometry, and interband coherence. In general, as the band indices are not equal, $\frac{ f_{0, {\bf k} }^{ m^{ \prime } } - f_{0, {\bf k} }^{m } }{ \varepsilon_{ {\bf k} }^{ m^{\prime} } - \varepsilon_{ {\bf k} }^{ m } }$ cannot be replaced by $\frac{ \partial f_{ 0, {\bf k } }^{ m } }{ \partial \varepsilon_{ {\bf k } }^{ m } }$. However, with a moderate temperature $T$ in the semiclassical regime satisfying $\vert \varepsilon_{ {\bf k} }^{ m } - \mu \vert \lesssim k_{ \mathrm{ B } } T$, $\frac{ f_{0, {\bf k} }^{ m^{ \prime } } - f_{0, {\bf k} }^{m } }{ \varepsilon_{ {\bf k} }^{ m^{\prime} } - \varepsilon_{ {\bf k} }^{ m } }$ behaves like a constant, and it can be approximated by $\frac{ \partial f_{ 0, {\bf k } }^{ m } }{ \partial \varepsilon_{ {\bf k } }^{ m } }$ evaluated at the Fermi surface, which we write as $\frac{ \partial f_{ 0 } }{ \partial \varepsilon }$ in the following for brevity. Further, we employ the commutation relation and the identity in Eq.~(\ref{eq:commutator-r-U}) and find that 
\begin{align}
\label{eq:J-S-4}
J_{ \mathrm{S}_{1}, {\bf k} }^{ m m } \left( t \right) =  e {\boldsymbol{\mathcal{E}}} (t) \cdot \left( {\bf v}_{ {\bf k} }^{  \mathrm{sj}, m } + {\bf v}_{ {\bf k} }^{  \mathrm{L}, m } \right) \frac{ \partial f_{ 0 } }{ \partial \varepsilon }.
\end{align} 
The term involving ${\bf v}_{ {\bf k} }^{  \mathrm{sj}, m }$ agrees with semiclassical wave packet analysis~\cite{Sinitsyn2006, Sinitsyn2007}. The other term containing the longitudinal velocity ${\bf v}_{ {\bf k} }^{  \mathrm{L}, m }$ defined in Eq.~(\ref{eq:v-sj-L}) is a quantum correction. $J_{ \mathrm{S}_{1}, {\bf k} }^{ m m } \left( t \right)$ is interpreted as a correction to the collision integral induced by the side-jump velocity ${\bf v}_{ {\bf k} }^{  \mathrm{sj}, m }$ and the longitudinal velocity ${\bf v}_{ {\bf k} }^{  \mathrm{L}, m }$. In the presence of an applied electric field, there is an additional displacement shift generated during scattering processes due to ${\bf v}_{ {\bf k} }^{  \mathrm{sj}, m }$ and ${\bf v}_{ {\bf k} }^{  \mathrm{L}, m }$. With the displacement shift, there is work done by the electric field, and the corresponding power is $- e {\boldsymbol{\mathcal{E}}} (t) \cdot \left( {\bf v}_{ {\bf k} }^{  \mathrm{sj}, m } + {\bf v}_{ {\bf k} }^{  \mathrm{L}, m } \right)$. This power in turn introduces corrections to the collision integral. If we neglect the displacement shift, elastic scattering from ${\bf k}$ to ${\bf k}^{ \prime }$ requires $\varepsilon_{ {\bf k} }^{ m } = \varepsilon_{ {\bf k}^{ \prime} }^{ m }$ demanded by the conservation of energy. However, when there is extra work done during scattering, electrons can be scattered from ${\bf k}$ to ${\bf k}^{ \prime }$ (or vice versa) even if $\varepsilon_{ {\bf k} }^{ m } \ne \varepsilon_{ {\bf k}^{ \prime} }^{ m }$, as long as the extra work compensates for the energy difference. The leading order correction to the collision integral is the product of the power and $\frac{ \partial f_{ 0 } }{ \partial \varepsilon }$, i.e., $J_{ \mathrm{S}_{1}, {\bf k} }^{ m m } \left( t \right)$. 

The leading-order contribution of ${\bf v}_{ {\bf k} }^{  \mathrm{L}, m }$ here is also longitudinal. Therefore, when it comes to Hall current, we may neglect ${\bf v}_{ {\bf k} }^{  \mathrm{L}, m }$ and write the following collision integral for $f_{0, {\bf k} }^{m }$~\cite{Sinitsyn2006},
\begin{align}
I_{0, {\bf k} }^{ m m } \left( t \right) = & \frac{2 \pi}{\hbar} n_{i} \int d{\bf k}^{\prime} V_{ {\bf k} {\bf k}^{ \prime } }^{ m m } V_{ {\bf k}^{ \prime} {\bf k} }^{ m m } \left( f_{ 0, {\bf k} }^{m} - f_{ 0, {\bf k}^{\prime} }^{m} \right) \nonumber \\
& \delta \left( \varepsilon_{ {\bf k}^{ \prime} }^{ m } - \varepsilon_{ {\bf k} }^{ m } + e {\boldsymbol{\mathcal{E}}} \left(t \right) \cdot \delta {\bf r}_{ {\bf k}^{ \prime } {\bf k} }^{m} \right),
\end{align}
where the side-jump correction is reflected in the energy conservation constraint in the $\delta$ function. We find that in the absence of ${\boldsymbol{\mathcal{E}}} \left(t \right)$ or without side jump, the collision integral above vanishes, as $f_{ 0, {\bf k} }^{m}$ describes the equilibrium, while the collision integral does not vanish if we apply ${\boldsymbol{\mathcal{E}}} \left(t \right)$ and include the side-jump position shift $\delta {\bf r}_{ {\bf k}^{ \prime } {\bf k} }^{m}$; i.e., the equilibrium distribution is no longer stable when side jump is introduced. 

Although the similarity between the side-jump contribution in the wave packet formalism and the corresponding correction in the density matrix was noticed a long time ago~\cite{Luttinger1958,Sinitsyn2006,Culcer2017,Atencia2022}, a thorough comparison of the two has not been made. Here, we show that at moderate temperatures the semiclassical side-jump contribution agrees with corresponding terms from the density matrix approach. Moreover, we identify a longitudinal velocity ${\bf v}_{ {\bf k} }^{  \mathrm{L}, m }$ which comes from the interband virtual transition and is missing in the semiclassical formalism. We also note that the semiclassical calculation of side-jump responses usually directly uses $\frac{ \partial f_{ 0, {\bf k } }^{ m } }{ \partial \varepsilon_{ {\bf k } }^{ m } }$ instead of $\frac{ f_{0, {\bf k} }^{ m^{ \prime } } - f_{0, {\bf k} }^{m } }{ \varepsilon_{ {\bf k} }^{ m^{\prime} } - \varepsilon_{ {\bf k} }^{ m } }$, sometimes even at $T=0 \ \mathrm{K}$. The approximation works when $\vert \varepsilon_{ {\bf k} }^{ m } - \mu \vert \lesssim k_{ \mathrm{ B } } T$ is satisfied, and at low temperatures it deviates from the result for the density matrix. 

Here we focus on side jumps arising from the band structure. An extrinsic side jump due to spin-orbit coupling in impurity potentials also exists, and the corresponding collision integral has been derived with the density matrix in Ref.~\cite{PhysRevB.81.125332}.

\section{Skew scattering}
Whereas the side-jump correction is included in the first Born approximation, a discussion of skew scattering requires going beyond the first order. Skew scattering refers to processes whose contributions to the scattering rate from ${\bf k}$ to ${\bf k}^{\prime}$ and from ${\bf k}^{\prime}$ to ${\bf k}$ have the same magnitude but opposite signs. The approximation leading to Eq.~(\ref{eq:Liouville-4}) discards many terms related to skew scattering in the quantum Liouville equation, Eq.~(\ref{eq:Liouville-2}). This can be seen if we plug Eq.~(\ref{eq:rho-integral}) into the commutator in Eq.~(\ref{eq:Liouville-2}) iteratively, to the $N$th order of $\hat{U}$,
\begin{align}
\label{eq:Liouville-N}
&\frac{d}{dt} \hat{\rho}_{E, \mathrm{I}}(t) - \left( - \frac{i}{\hbar} \right)
\left[ \hat{H}_{E, I} (t), \hat{\rho}_{0} \right] \nonumber \\
- & \left( - \frac{i}{\hbar} \right)^{2} \int_{-\infty}^{t} d t_{1} \left[ \hat{U}_{\mathrm{I}}(t), \left[ \hat{H}_{E, I} (t_{1}), \hat{\rho}_{0} \right] \right] \nonumber \\
- & \left( - \frac{i}{\hbar} \right)^{3} \int_{-\infty}^{t} d t_{1} \int_{-\infty}^{ t_{1} } d t_{2} \left[ \hat{U}_{\mathrm{I}}(t),  \left[ \hat{U}_{\mathrm{I}}(t_{1}), \left[ \hat{H}_{E, I} (t_{2}), \hat{\rho}_{0} \right] \right] \right] \nonumber \\
- & \cdots \nonumber \\
- & \left( - \frac{i}{\hbar} \right)^{N} \int_{-\infty}^{t} d t_{1} \int_{-\infty}^{ t_{1} } d t_{2} \cdots \int_{-\infty}^{ t_{N-2} } d t_{N-1} \nonumber \\
&\left[ \hat{U}_{\mathrm{I}}(t),  \left[ \hat{U}_{\mathrm{I}}(t_{1}), \left[\hat{U}_{\mathrm{I}}(t_{2}), \cdots \left[ \hat{H}_{E, I} (t_{N - 1}), \hat{\rho}_{0} \right] \cdots \right] \right] \right] \nonumber \\
- & \left( - \frac{i}{\hbar} \right)^{N} \int_{-\infty}^{t} d t_{1} \int_{-\infty}^{ t_{1} } d t_{2} \cdots \int_{-\infty}^{ t_{N-2} } d t_{N-1} \nonumber \\
&\left[ \hat{U}_{\mathrm{I}}(t),  \left[ \hat{U}_{\mathrm{I}}(t_{1}), \left[\hat{U}_{\mathrm{I}}(t_{2}), \cdots \left[ \hat{U}_{\mathrm{I}}(t_{N - 1}), \hat{\rho}_{E, \mathrm{I}}(t_{N - 1}) \right] \cdots \right] \right] \right] \nonumber \\
 = &  0.
\end{align}
For comparison, the same procedure for Eq.~(\ref{eq:Liouville-4}) produces only terms with even numbers of $\hat{U}_{I}$'s in Eq.~(\ref{eq:Liouville-N}), and the impurity average is taken pairwise with the first Born approximation. Terms with odd numbers of $\hat{U}_{I}$'s are omitted in Eq.~(\ref{eq:Liouville-4}). Hence, the first-Born approximation completely misses the skew scattering at order $V^{3}$.

To recover the skew-scattering terms at order $V^{3}$, we go one order higher than the first-Born approximation and write Eq.~(\ref{eq:Liouville-N}) to the third order of $\hat{U}_{I}$,
\begin{align}
\label{eq:Liouville-5}
&\frac{d}{dt} \hat{\rho}_{E, \mathrm{I}}(t) - \left( - \frac{i}{\hbar} \right)
\left[ \hat{H}_{E, I} (t), \hat{\rho}_{0} \right] \nonumber \\
- & \left( - \frac{i}{\hbar} \right)^{2} \int_{-\infty}^{t} d t_{1} \left[ \hat{U}_{\mathrm{I}}(t), \left[ \hat{H}_{E, I} (t_{1}), \hat{\rho}_{0} \right] \right] \nonumber \\
- & \left( - \frac{i}{\hbar} \right)^{3} \int_{-\infty}^{t} d t_{1} \int_{-\infty}^{ t_{1} } d t_{2} \left[ \hat{U}_{\mathrm{I}}(t),  \left[ \hat{U}_{\mathrm{I}}(t_{1}), \left[ \hat{H}_{E, I} (t_{2}), \hat{\rho}_{0} \right] \right] \right] \nonumber \\
- & \left( - \frac{i}{\hbar} \right)^{3} \int_{-\infty}^{t} d t_{1} \int_{-\infty}^{ t_{1} } d t_{2} \nonumber \\
& \left[ \hat{U}_{\mathrm{I}}(t),  \left[ \hat{U}_{\mathrm{I}}(t_{1}), \left[\hat{U}_{\mathrm{I}}(t_{2}), \hat{\rho}_{E, \mathrm{I}}(t_{2}) \right] \right] \right] \nonumber \\
 = &  0.
\end{align}
The first two terms already appear in Eq.~(\ref{eq:Liouville-4}), and the third term vanishes after we take the impurity average. The fourth term is contained in Eq.~(\ref{eq:Liouville-4}), as we can check by substituting the integral form of $\hat{\rho}_{E, \mathrm{I}}(t)$ in Eq.~(\ref{eq:Liouville-4}) into the commutator in Eq.~(\ref{eq:J-t}). The last term is the source of the skew-scatteing contribution at order $V^{3}$. We now keep terms one order higher than the first Born approximation and add the last term to the quantum Liouville equation of Eq.~(\ref{eq:Liouville-4}), 
\begin{align}
\label{eq:Liouville-6}
\frac{d}{dt} \hat{\rho}_{E, \mathrm{I}}(t) + \frac{i}{\hbar} 
\left[ \hat{H}_{E, I} (t), \hat{\rho}_{0} \right] + \mathcal{J} \left( t \right) + \mathcal{J}_{ 3 } \left( t \right) = 0,
\end{align}
\begin{align}
\label{eq:J-sk-0}
& \mathcal{J}_{ 3 } \left( t \right) =  - \frac{ i }{ \hbar^{3} } \nonumber \\
& \times \int_{-\infty}^{t} d t_{1} \int_{-\infty}^{ t_{1} } d t_{2}  \left[ \hat{U}_{\mathrm{I}}(t),  \left[ \hat{U}_{\mathrm{I}}(t_{1}), \left[\hat{U}_{\mathrm{I}}(t_{2}), \hat{\rho}_{E, \mathrm{I}}(t_{2}) \right] \right] \right].
\end{align}
We follow the same procedure to analyze the correction to the semiclassical distribution function, namely, the diagonal components of the quantum density matrix induced by skew scattering, and we evaluate the diagonal components of $\mathcal{J}_{ 3 } \left( t \right)$, $J_{3, {\bf k} }^{ m m } \left( t \right) = \langle m, {\bf k}, t \vert \mathcal{J}_{ 3 } \left( t \right) \vert m, {\bf k}, t \rangle$. According to the analysis in Sec.~\ref{sec:Boltzmann}, the diagonal components of the density matrix $f_{ {\bf k} }^{m} \left( t \right)$ have a lower order than its off-diagonal counterpart $S_{ {\bf k} }^{ m m^{ \prime } } \left( t \right)$ (see Table~\ref{Table:Order}). Here, we focus on the part of $J_{ 3, {\bf k} }^{ m m } \left( t \right)$ induced by $f_{ {\bf k} }^{m} \left( t \right)$, $J_{ 3, f, {\bf k} }^{ m m }$, which satisfies 
\begin{align}
\label{eq:J-sk-mm-2} 
J_{3, f, {\bf k} }^{ m m } \left( t \right) = & \sum_{ m^{\prime} } \int d {\bf k}^{\prime} \left[ w^{3, m^{\prime} m }_{ {\bf k}^{\prime} {\bf k} } f_{ {\bf k} }^{m} \left( t \right) - w^{3, m m^{\prime} }_{ {\bf k} {\bf k}^{\prime} } f_{ {\bf k}^{\prime} }^{ m^{\prime} } \left( t \right) \right], \\
\label{eq:w-3} 
w^{3, m^{\prime} m }_{ {\bf k}^{\prime} {\bf k} } = & \frac{ 4 \pi }{ \hbar } \delta \left( \varepsilon_{ {\bf k} }^{ m } - \varepsilon_{ {\bf k}^{ \prime} }^{ m^{ \prime} } \right) \sum_{ m^{ \prime \prime } } \int d {\bf k}^{ \prime \prime }  \nonumber \\
& \times \bigg[ \frac{ \mathcal{V. P.} }{ \varepsilon_{ {\bf k} }^{m} - \varepsilon_{ {\bf k}^{ \prime \prime } }^{ m^{ \prime \prime } }} \mathrm{Re} \left( U_{ {\bf k}^{ \prime } {\bf k} }^{ m^{ \prime} m } U_{ {\bf k} {\bf k}^{ \prime \prime } }^{ m m^{ \prime \prime } } U_{ {\bf k}^{ \prime \prime } {\bf k}^{ \prime} }^{ m^{ \prime \prime } m^{ \prime} } \right) \nonumber \\
& - \pi \delta \left( \varepsilon_{ {\bf k} }^{ m } - \varepsilon_{ {\bf k}^{ \prime \prime } }^{ m^{ \prime \prime } } \right) \mathrm{Im} \left( U_{ {\bf k}^{ \prime } {\bf k} }^{ m^{ \prime} m } U_{ {\bf k} {\bf k}^{ \prime \prime } }^{ m m^{ \prime \prime } } U_{ {\bf k}^{ \prime \prime } {\bf k}^{ \prime} }^{ m^{ \prime \prime } m^{ \prime} } \right) \bigg] .
\end{align}
$J_{3, f, {\bf k} }^{ m m } \left( t \right)$ is in the form of a collision integral with $w^{3, m^{\prime} m }_{ {\bf k}^{\prime} {\bf k} }$ being the scattering rate. The first term in the square brackets in Eq.~(\ref{eq:w-3}) renormalizes the symmetric scattering rate, and it is usually discarded. The second term describes skew-scattering at order $V^{3}$. We extract the skew-scattering rate from Eq.~(\ref{eq:w-3}), $w^{3\mathrm{A}, m }_{ {\bf k} {\bf k}^{\prime} } \equiv \frac{1}{2} \left( w^{3, m m }_{ {\bf k} {\bf k}^{\prime} } - w^{3, m  m }_{ {\bf k}^{\prime} {\bf k} } \right)$,
\begin{align}
\label{eq:w-A-m}
w^{ 3\mathrm{A}, m }_{ {\bf k} {\bf k}^{\prime} } = & - \frac{ 4 \pi^{2} }{ \hbar } n_{i} \int d{\bf k}^{ \prime \prime }  \delta \left( \varepsilon_{ {\bf k} }^{ m } - \varepsilon_{ {\bf k}^{ \prime} }^{ m } \right) \delta \left( \varepsilon_{ {\bf k} }^{ m } - \varepsilon_{ {\bf k}^{ \prime \prime } }^{ m } \right) \nonumber \\
& \mathrm{Im} \left( V_{ {\bf k} {\bf k}^{ \prime } }^{ m m } V_{ {\bf k}^{ \prime} {\bf k}^{ \prime \prime } }^{ m m } V_{ {\bf k}^{ \prime \prime } {\bf k} }^{ m m } \right),
\end{align}
where we have utilized the assumption that the bands are separated, and we have performed an ensemble average over the impurity configurations (see Appendix \ref{sec:disorder-averaging} for details). We can easily check that $w^{ 3 \mathrm{A}, m }_{ {\bf k} {\bf k}^{\prime} }$ is antisymmetric, i.e., $w^{ 3 \mathrm{A}, m }_{ {\bf k} {\bf k}^{\prime} } = - w^{ 3 \mathrm{A}, m }_{ {\bf k}^{\prime} {\bf k} }$. It has been shown that $w^{ 3 \mathrm{A}, m }_{ {\bf k} {\bf k}^{\prime} }$ and the side-jump position shift $\delta {\bf r}_{ {\bf k}^{ \prime } {\bf k} }^{m}$ originate from the same gauge-invariant geometric phase~\cite{Sinitsyn2006}. 

The Gaussian skew-scattering contribution at order $V^{4}$ can be obtained in a similar fashion. First we add the next-order term $\mathcal{J}_{ 4 } \left( t \right)$ to the quantum Liouville equation,
\begin{align}
\label{eq:J-4}
& \mathcal{J}_{ 4 } \left( t \right) =  - \frac{ 1 }{ \hbar^{ 4 } } \int_{-\infty}^{t} d t_{1} \int_{-\infty}^{ t_{1} } d t_{2} \int_{-\infty}^{ t_{ 2 } } d t_{ 3 } \nonumber \\
& \left[ \hat{U}_{\mathrm{I}}(t),  \left[ \hat{U}_{\mathrm{I}}(t_{1}), \left[\hat{U}_{\mathrm{I}}(t_{2}), \left[\hat{U}_{\mathrm{I}}(t_{3}), \hat{\rho}_{E, \mathrm{I}}(t_{3}) \right] \right] \right] \right].
\end{align}
Again, we focus on the diagonal component of $\mathcal{J}_{ 4 } \left( t \right)$, $J_{4, {\bf k} }^{ m m } \left( t \right) = \langle m, {\bf k}, t \vert \mathcal{J}_{ 4 } \left( t \right) \vert m, {\bf k}, t \rangle$, as it is more important than the off-diagonal components. Further, the leading-order contribution to $J_{4, {\bf k} }^{ m m } \left( t \right)$ is induced by the diagonal $f_{ {\bf k} }^{m} \left( t \right)$ instead of the off-diagonal $S_{ {\bf k} }^{ m m^{ \prime } } \left( t \right)$. We denote this part as $J_{4, f, {\bf k} }^{ m m } \left( t \right)$, and a straightforward calculation yields
\begin{align}
J_{4, f, {\bf k} }^{ m m } \left( t \right) = & \sum_{ m^{\prime} } \int d {\bf k}^{\prime} \left[ w^{4, m^{\prime} m }_{ {\bf k}^{\prime} {\bf k} } f_{ {\bf k} }^{m} \left( t \right) - w^{4, m m^{\prime} }_{ {\bf k} {\bf k}^{\prime} } f_{ {\bf k}^{\prime} }^{ m^{\prime} } \left( t \right) \right], \\
w^{4, m^{\prime} m }_{ {\bf k}^{\prime} {\bf k} } = & \frac{ 2 \pi }{ \hbar } \delta \left( \varepsilon_{ {\bf k} }^{ m } - \varepsilon_{ {\bf k}^{ \prime} }^{ m^{ \prime} } \right) \sum_{ m_{1}, m_{2} } \int d {\bf k}_{1} \int d {\bf k}_{2} \nonumber \\
& \times \bigg[ \frac{ U_{ {\bf k} {\bf k}^{ \prime } }^{ m m^{ \prime } } U_{ {\bf k}^{ \prime } {\bf k}_{1} }^{ m^{ \prime } m_{1} } }{ \varepsilon_{ {\bf k} }^{ m }- \varepsilon_{ {\bf k}_{1} }^{ m_{1} } + i \hbar \eta } \frac{ U_{ {\bf k}_{1} {\bf k}_{2} }^{ m_{1} m_{2} } U_{ {\bf k}_{2} {\bf k}_{1} }^{ m_{ 2 } m } }{ \varepsilon_{ {\bf k} }^{ m }- \varepsilon_{ {\bf k}_{ 2 } }^{ m_{ 2 } } + i \hbar \eta } \nonumber \\
& + \frac{ U_{ {\bf k} {\bf k}_{1} }^{ m m_{1} } U_{ {\bf k}_{1} {\bf k}^{ \prime } }^{ m_{1} m^{ \prime } } }{ \varepsilon_{ {\bf k} }^{ m }- \varepsilon_{ {\bf k}_{1} }^{ m_{1} } - i \hbar \eta } \frac{ U_{ {\bf k}^{ \prime } {\bf k}_{2} }^{ m^{ \prime } m_{2} } U_{ {\bf k}_{2} {\bf k} }^{ m_{ 2 } m } }{ \varepsilon_{ {\bf k} }^{ m }- \varepsilon_{ {\bf k}_{ 2 } }^{ m_{ 2 } } + i \hbar \eta } \nonumber \\
& + \frac{ U_{ {\bf k} {\bf k}_{ 2 } }^{ m m_{ 2 } } U_{ {\bf k}_{ 2 } {\bf k}_{1} }^{ m_{ 2 } m_{1} } }{ \varepsilon_{ {\bf k} }^{ m }- \varepsilon_{ {\bf k}_{1} }^{ m_{1} } - i \hbar \eta } \frac{ U_{ {\bf k}_{1} {\bf k}^{ \prime } }^{ m_{1} m^{ \prime } } U_{ {\bf k}^{ \prime } {\bf k}_{1} }^{ m^{ \prime } m } }{ \varepsilon_{ {\bf k} }^{ m }- \varepsilon_{ {\bf k}_{ 2 } }^{ m_{ 2 } } - i \hbar \eta } \bigg] .
\end{align}
The scattering rate $w^{4, m^{\prime} m }_{ {\bf k}^{\prime} {\bf k} }$ contains a symmetric part and an antisymmetric part. Similar to the case for $J_{ 3, f, {\bf k} }^{ m m } \left( t \right)$, the symmetric part corresponds to trivial renormalization of conventional scattering rate, while the antisymmetric part represents Gaussian skew scattering. Different from $J_{ 3, f, {\bf k} }^{ m m } \left( t \right)$, part of $J_{4, f, {\bf k} }^{ m m } \left( t \right)$ is already present in the first Born approximation in the previous sections, for example, the contribution of $S_{ 2, {\bf k} }^{ m m } \left( t \right)$ to $J_{ S, {\bf k} }^{ m m } \left( t \right)$. To avoid double counting, we exclude it in Sec.~\ref{sec:side-jump}. Averaging over impurity configurations (See Appendix \ref{sec:disorder-averaging} for details), we have the Gaussian skew scattering rate $w^{4\mathrm{A}, m }_{ {\bf k} {\bf k}^{\prime} } \equiv \frac{1}{2} \left( w^{4, m m }_{ {\bf k} {\bf k}^{\prime} } - w^{4, m  m }_{ {\bf k}^{\prime} {\bf k} } \right)$, 
\begin{align}
w^{4\mathrm{A}, m }_{ {\bf k} {\bf k}^{\prime} } = & \sum_{i=1}^{5} w^{4\mathrm{A}-i, m }_{ {\bf k} {\bf k}^{\prime} }, \\
w^{4\mathrm{A}-1, m }_{ {\bf k} {\bf k}^{\prime} } = & \frac{ 4 \pi }{ \hbar } n_{i}^{2} \delta \left( \varepsilon_{ {\bf k} }^{ m } - \varepsilon_{ {\bf k}^{ \prime} }^{ m } \right) \sum_{ m_{1}, m_{2} } \int d {\bf k}_{1} \nonumber \\
& \times \mathrm{Im} \left( \frac{ 1 }{ \varepsilon_{ {\bf k} }^{ m }- \varepsilon_{ {\bf k}_{1} }^{ m_{1} } - i \hbar \eta } \frac{ 1 }{ \varepsilon_{ {\bf k} }^{ m }- \varepsilon_{ {\bf k}^{ \prime } }^{ m_{ 2 } } - i \hbar \eta } \right) \nonumber \\
& \times \mathrm{Im} \left( V_{ {\bf k} {\bf k}^{ \prime } }^{ m m } V_{ {\bf k}^{ \prime } {\bf k}_{1} }^{ m m_{1} } V_{ {\bf k}_{1} {\bf k}^{ \prime } }^{ m_{1} m_{2} } V_{ {\bf k}^{ \prime } {\bf k} }^{ m_{ 2 } m } \right), \\
w^{4\mathrm{A}-2, m }_{ {\bf k} {\bf k}^{\prime} } = & \frac{ 4 \pi }{ \hbar } n_{i}^{2} \delta \left( \varepsilon_{ {\bf k} }^{ m } - \varepsilon_{ {\bf k}^{ \prime} }^{ m } \right) \sum_{ m_{1}, m_{2} } \int d {\bf k}_{1} \nonumber \\
& \times \mathrm{Im} \left( \frac{ 1 }{ \varepsilon_{ {\bf k} }^{ m }- \varepsilon_{ {\bf k} }^{ m_{1} } - i \hbar \eta } \frac{ 1 }{ \varepsilon_{ {\bf k} }^{ m }- \varepsilon_{ {\bf k}_{1} }^{ m_{ 2 } } - i \hbar \eta } \right) \nonumber \\
& \times \mathrm{Im} \left( V_{ {\bf k} {\bf k}^{ \prime } }^{ m m } V_{ {\bf k}^{ \prime } {\bf k} }^{ m m_{1} } V_{ {\bf k} {\bf k}_{1} }^{ m_{1} m_{2} } V_{ {\bf k}_{1} {\bf k} }^{ m_{ 2 } m } \right), \\
w^{4\mathrm{A}-3, m }_{ {\bf k} {\bf k}^{\prime} } = & \frac{ 2 \pi }{ \hbar } n_{i}^{2} \delta \left( \varepsilon_{ {\bf k} }^{ m } - \varepsilon_{ {\bf k}^{ \prime} }^{ m } \right) \sum_{ m_{1}, m_{2} } \int d {\bf k}_{1} \nonumber \\
& \times \mathrm{Im} \left( \frac{ 1 }{ \varepsilon_{ {\bf k} }^{ m }- \varepsilon_{ {\bf k}_{1} }^{ m_{1} } + i \hbar \eta } \frac{ 1 }{ \varepsilon_{ {\bf k} }^{ m }- \varepsilon_{ {\bf k}_{1} }^{ m_{ 2 } } - i \hbar \eta } \right) \nonumber \\
& \times \mathrm{Im} \left( V_{ {\bf k} {\bf k}_{1} }^{ m m_{1} } V_{ {\bf k}_{1} {\bf k}^{\prime} }^{ m_{1} m } V_{ {\bf k}^{\prime} {\bf k}_{1} }^{ m m_{2} } V_{ {\bf k}_{1} {\bf k} }^{ m_{ 2 } m } \right), \\
w^{4\mathrm{A}-4, m }_{ {\bf k} {\bf k}^{\prime} } = & \frac{ 4 \pi }{ \hbar } n_{i}^{2} \delta \left( \varepsilon_{ {\bf k} }^{ m } - \varepsilon_{ {\bf k}^{ \prime} }^{ m } \right) \sum_{ m_{1}, m_{2} } \int d {\bf k}_{1} \nonumber \\
& \times \mathrm{Im} \left( \frac{ 1 }{ \varepsilon_{ {\bf k} }^{ m }- \varepsilon_{ {\bf k}_{1} }^{ m_{1} } - i \hbar \eta } \frac{ 1 }{ \varepsilon_{ {\bf k} }^{ m }- \varepsilon_{ {\bf k} - {\bf k}^{ \prime } + {\bf k}_{1} }^{ m_{ 2 } } - i \hbar \eta } \right) \nonumber \\
& \times \mathrm{Im} \left( V_{ {\bf k} {\bf k}^{ \prime } }^{ m m } V_{ {\bf k}^{ \prime } {\bf k}_{1} }^{ m m_{1} } V_{ {\bf k}_{1}, {\bf k} - {\bf k}^{ \prime } + {\bf k}_{1} }^{ m_{1} m_{2} } V_{ {\bf k} - {\bf k}^{ \prime } + {\bf k}_{1}, {\bf k} }^{ m_{ 2 } m } \right), \\
w^{4\mathrm{A}-5, m }_{ {\bf k} {\bf k}^{\prime} } = & \frac{ 2 \pi }{ \hbar } n_{i}^{2} \delta \left( \varepsilon_{ {\bf k} }^{ m } - \varepsilon_{ {\bf k}^{ \prime} }^{ m } \right) \sum_{ m_{1}, m_{2} } \int d {\bf k}_{1} \nonumber \\
& \times \mathrm{Im} \left( \frac{ 1 }{ \varepsilon_{ {\bf k} }^{ m }- \varepsilon_{ {\bf k}_{1} }^{ m_{1} } + i \hbar \eta } \frac{ 1 }{ \varepsilon_{ {\bf k} }^{ m }- \varepsilon_{ {\bf k} - {\bf k}^{ \prime } + {\bf k}_{1} }^{ m_{ 2 } } - i \hbar \eta } \right) \nonumber \\
& \times \mathrm{Im} \left( V_{ {\bf k} {\bf k}_{1} }^{ m m_{1} } V_{ {\bf k}_{1} {\bf k}^{ \prime } }^{ m_{1} m } V_{ {\bf k}^{ \prime } , {\bf k} - {\bf k}^{ \prime } + {\bf k}_{1} }^{ m m_{2} } V_{ {\bf k} - {\bf k}^{ \prime } + {\bf k}_{1}, {\bf k} }^{ m_{ 2 } m } \right).
\end{align} 
The expression for $w^{4\mathrm{A}, m }_{ {\bf k} {\bf k}^{\prime} }$ indicates that the Gaussian skew scattering inevitably involves an interband virtual transition. In comparison, skew scattering at order $V^{3}$ happens within one single band. As noted in earlier works~\cite{Sinitsyn2006,Sinitsyn2007PRB,Nagaosa2010}, scaling of the Hall current induced by Gaussian skew scattering is similar to that induced by a side jump. Previous works~\cite{Kohn-Luttinger1957,Luttinger1958,Culcer2017,Atencia2022} did not fully perform the ensemble average over different impurity configurations, so their Gaussian skew scattering rates were incomplete and disagree with results derived with other approaches. By comparison, here we have the full Gaussian skew scattering rate, including components that appear in the non-crossing approximation ($w^{4\mathrm{A}-1, m }_{ {\bf k} {\bf k}^{\prime} }$, $w^{4\mathrm{A}-2, m }_{ {\bf k} {\bf k}^{\prime} }$ and $w^{4\mathrm{A}-3, m }_{ {\bf k} {\bf k}^{\prime} }$) and the crossing terms ($w^{4\mathrm{A}-4, m }_{ {\bf k} {\bf k}^{\prime} }$ and $w^{4\mathrm{A}-5, m }_{ {\bf k} {\bf k}^{\prime} }$).

We combine the skew-scattering parts of $J_{3, f, {\bf k} }^{ m m } \left( t \right)$ and $J_{4, f, {\bf k} }^{ m m } \left( t \right)$ as $J_{ \mathrm{sk}, {\bf k} }^{ m m } \left( t \right)$, 
\begin{align}
\label{eq:J-sk}
J_{ \mathrm{sk}, {\bf k} }^{ m m } \left( t \right) = \int d {\bf k}^{\prime} \left[ w^{\mathrm{A}, m }_{ {\bf k}^{\prime} {\bf k} } f_{ {\bf k} }^{m} \left( t \right) - w^{\mathrm{A}, m }_{ {\bf k} {\bf k}^{\prime} } f_{ {\bf k}^{\prime} }^{ m } \left( t \right) \right],
\end{align} 
where the skew scattering rate $w^{\mathrm{A}, m }_{ {\bf k}^{\prime} {\bf k} }$ is the sum of the skew-scattering rate at order $V^{3}$ and the Gaussian skew scattering rate, $w^{\mathrm{A}, m }_{ {\bf k}^{\prime} {\bf k} } = w^{\mathrm{3A}, m }_{ {\bf k}^{\prime} {\bf k} } + w^{\mathrm{4A}, m }_{ {\bf k}^{\prime} {\bf k} }$. It is straightforward to see that $\int d{\bf k}^{\prime} w^{ \mathrm{A}, m }_{ {\bf k}^{\prime} {\bf k} } = 0$~\cite{Konig2019,Liang2020,Ma2023}, so that $J_{ \mathrm{sk}, {\bf k} }^{ m m } \left( t \right)$ can be simplified as $J_{ \mathrm{sk}, {\bf k} }^{ m m } \left( t \right) = - \int d{\bf k}^{\prime} w^{ \mathrm{A}, m }_{ {\bf k} {\bf k}^{\prime} } f_{ {\bf k}^{\prime} }^{m} \left( t \right)$. 

Following our analysis, $J_{ \mathrm{sk}, {\bf k} }^{ m m } \left( t \right)$ will appear in the revised diagonal component of the quantum Liouville equation of Eq.~(\ref{eq:Liouville-6}),
\begin{align}
\label{eq:diagonal-with-skew}
&\frac{d}{dt} f_{ {\bf k} }^{m} \left( t \right) - \frac{e}{\hbar} {\boldsymbol{\mathcal{E}}}(t) \cdot \nabla_{ {\bf k} } f_{0, {\bf k} }^{m } + J_{ f, {\bf k} }^{ m m } \left( t \right) + J_{ S, {\bf k} }^{ m m } \left( t \right)  \nonumber \\
+ & J_{ \mathrm{sk}, {\bf k} }^{ m m } \left( t \right)  = 0. 
\end{align}
Like in Eq.~(\ref{eq:diagonal-1}), inclusion of $J_{ \mathrm{sk}, {\bf k} }^{ m m } \left( t \right)$ results in a correction to the distribution function $f_{ {\bf k} }^{m} \left( t \right)$, and this term is denoted as $f_{ \mathrm{sk}, {\bf k} }^{m} \left( t \right)$. Equation~(\ref{eq:f_k-f_1-f_sj}) is now replaced by
\begin{equation}
f_{ {\bf k} }^{m} \left( t \right) = f_{ 1, {\bf k} }^{m} \left( t \right) + f_{ \mathrm{sj}, {\bf k} }^{m} \left( t \right) + f_{ \mathrm{sk}, {\bf k} }^{m} \left( t \right),
\end{equation}
where the conventional part $f_{ 1, {\bf k} }^{m} \left( t \right)$ and the side-jump part $f_{ \mathrm{sj}, {\bf k} }^{m} \left( t \right)$ are given by Eqs.~(\ref{eq:diagonal-2}) and (\ref{eq:side-jump-diagonal}), respectively. With the expression for $J_{ f, {\bf k} }^{ m m } \left( t \right)$ in Eq.~(\ref{eq:J-f-k}), we see that $f_{ \mathrm{sk}, {\bf k} }^{m} \left( t \right)$ gives rise to an additional term $I_{ \mathrm{sk}, {\bf k} }^{ m } \left( t \right)$ in the collision integral,
\begin{equation}
J_{ f, {\bf k} }^{ m m } \left( t \right) = I_{ 1, {\bf k} }^{ m } \left( t \right) + I_{ \mathrm{sj}, {\bf k} }^{ m } \left( t \right) + I_{ \mathrm{sk}, {\bf k} }^{ m } \left( t \right),
\end{equation}
where $I_{ 1, {\bf k} }^{ m } \left( t \right)$ and $I_{ \mathrm{sj}, {\bf k} }^{ m } \left( t \right)$ are defined in Eqs.~(\ref{eq:I-1}) and (\ref{eq:I-sj}), respectively. $I_{ \mathrm{sk}, {\bf k} }^{ m } \left( t \right)$ satisfies
\begin{equation}
\label{eq:I-sk}
I_{ \mathrm{sk}, {\bf k} }^{ m } \left( t \right) = \int d{\bf k}^{\prime} w^{ \mathrm{S}, m }_{ {\bf k} {\bf k}^{\prime} } \left[ f_{ \mathrm{sk}, {\bf k} }^{m} \left( t \right) - f_{ \mathrm{sk}, {\bf k}^{\prime} }^{m} \left( t \right) \right].
\end{equation} 
Now, if we extract the terms involving $f_{ \mathrm{sk}, {\bf k} }^{m} \left( t \right)$ in Eq.~(\ref{eq:diagonal-with-skew}), 
\begin{align}
\label{eq:skew-scattering-diagonal}
\frac{d}{dt}f_{ \mathrm{sk}, {\bf k} }^{m} \left( t \right) + I_{ \mathrm{sk}, {\bf k} }^{ m } \left( t \right) + J_{ \mathrm{sk}, {\bf k} }^{ m m } \left( t \right) = 0.
\end{align}
At leading order, we can use $J_{ \mathrm{sk}, {\bf k} }^{ m m } \left( t \right) = - \int d{\bf k}^{\prime} w^{ \mathrm{A}, m }_{ {\bf k} {\bf k}^{\prime} } f_{ 1, {\bf k}^{\prime} }^{m} \left( t \right)$ for Eq.~(\ref{eq:skew-scattering-diagonal}). 

If we take the skew-scattering contribution into account, the current density ${\bf j} \left( t \right)$ in Eq.~(\ref{eq:j-four-components}) is now replaced by 
\begin{equation}
\label{eq:j-5}
{\bf j} \left( t \right) =  {\bf j}_{1} \left( t \right) + {\bf j}_{\mathrm{sj-f}} \left( t \right) + {\bf j}_{\mathrm{a}} \left( t \right) + {\bf j}_{\mathrm{sj-v}} \left( t \right) + {\bf j}_{\mathrm{sk}} \left( t \right),
\end{equation}
where the conventional current ${\bf j}_{1} \left( t \right)$, the anomalous current ${\bf j}_{\mathrm{a}} \left( t \right)$, the velocity part of the side-jump current ${\bf j}_{\mathrm{sj-v}} \left( t \right)$ and the distribution function part of the side-jump current ${\bf j}_{\mathrm{sj-f}} \left( t \right)$ are provided in Eqs.~(\ref{eq:j-1}), (\ref{eq:j-a}), (\ref{eq:j-sj-v-3}) and (\ref{eq:j-sj-f}), respectively. The current induced by skew-scattering is
\begin{equation}
{\bf j}_{\mathrm{sk}} \left( t \right) = - e \sum_{ m } \int d {\bf k} \ {\bf v}_{ {\bf k} }^{ m } f_{ \mathrm{sk}, {\bf k} }^{m} \left( t \right) ,
\end{equation} 
combining the conventional group velocity ${\bf v}_{ {\bf k} }^{ m }$ and the skew-scattering correction to the distribution function $f_{ \mathrm{sk}, {\bf k} }^{m} \left( t \right)$. 

Actually, Eq.~(\ref{eq:diagonal-with-skew}) can be directly compared with the semiclassical Boltzmann equation with the skew-scattering and the side-jump corrections. To see this, we introduce $\bar{f}_{ {\bf k} }^{m} \left( t \right) = f_{ {\bf k} }^{m} \left( t \right) + f_{0, {\bf k} }^{ m }$, which is the distribution function in the semiclassical Boltzmann formalism, and the sum of the scattering terms corresponds to the collision intergral $\bar{I}_{ {\bf k} }^{m} \left( t \right)$, i.e., $\bar{I}_{ {\bf k} }^{m} \left( t \right) = - \left[ J_{ f, {\bf k} }^{ m m } \left( t \right) + J_{ S, {\bf k} }^{ m m } \left( t \right) + J_{ \mathrm{sk}, {\bf k} }^{ m m } \left( t \right) \right]$,
\begin{align}
\label{eq:full-collision-integral}
\bar{I}_{ {\bf k} }^{m} \left( t \right) = & - \int d{\bf k}^{\prime} \bigg[ w^{ m }_{ {\bf k}^{\prime} {\bf k} } \bar{f}_{ {\bf k} }^{m} \left( t \right) - w^{ m }_{ {\bf k} {\bf k}^{\prime} } \bar{f}_{ {\bf k}^{\prime} }^{m} \left( t \right) \bigg] \nonumber \\ 
& - e {\boldsymbol{\mathcal{E}}} (t) \cdot \left( {\bf v}_{ {\bf k} }^{  \mathrm{sj}, m } + {\bf v}_{ {\bf k} }^{  \mathrm{L}, m } \right) \frac{ \partial f_{ 0 } }{ \partial \varepsilon } ,
\end{align}
where $w^{ m }_{ {\bf k} {\bf k}^{\prime} } = w^{ \mathrm{S}, m }_{ {\bf k} {\bf k}^{\prime} } + w^{ \mathrm{A}, m }_{ {\bf k} {\bf k}^{\prime} }$ is the total scattering rate, including both symmetric (S) and antisymmetric (A) skew scattering. We can then rewrite Eq.~(\ref{eq:diagonal-with-skew}) into a compact form, 
\begin{align}
\label{eq:final-Boltzmann}
\frac{d}{dt} \bar{f}_{ {\bf k} }^{m} \left( t \right) - \frac{e}{\hbar} {\boldsymbol{\mathcal{E}}}(t) \cdot \nabla_{ {\bf k} } f_{0, {\bf k} }^{ m }  = \bar{I}_{ {\bf k} }^{m} \left( t \right) .
\end{align}
The charge current density ${\bf j} \left( t \right) $ can also be rewritten in a compact form as 
\begin{align}
{\bf j} \left( t \right) = & - e \sum_{ m } \int d {\bf k} \ \bar{\bf v}_{ {\bf k} }^{  m } \left( t  \right) \bar{f}_{ {\bf k} }^{m} \left( t \right) ,
\end{align}
where $\bar{\bf v}_{ {\bf k} }^{  m } \left( t  \right)$ is the total velocity, 
\begin{align}
\bar{\bf v}_{ {\bf k} }^{  m } \left( t  \right) = {\bf v}_{ {\bf k} }^{  m } + \frac{ e }{ \hbar }{\boldsymbol{\mathcal{E}}} \left( t \right) \times {\bf \Omega}^{ m }_{ {\bf k} } + {\bf v}_{ {\bf k} }^{  \mathrm{sj}, m } + {\bf v}_{ {\bf k} }^{  \mathrm{L}, m },
\end{align} 
comprising the conventional group velocity ${\bf v}_{ {\bf k} }^{  m }$, the anomalous velocity induced by the Berry curvature $\frac{ e }{ \hbar }{\boldsymbol{\mathcal{E}}} \left( t \right) \times {\bf \Omega}^{ m }_{ {\bf k} }$, the side-jump velocity ${\bf v}_{ {\bf k} }^{  \mathrm{sj}, m }$, and the longitudinal velocity ${\bf v}_{ {\bf k} }^{  \mathrm{L}, m }$. We see here that the semiclassical Boltzmann equation can be reproduced by quantum mechanical derivation at moderate temperatures, except that a correction, the longitudinal velocity ${\bf v}_{ {\bf k} }^{  \mathrm{L}, m }$, is introduced to the side-jump velocity. Note that here, we do not include terms combining skew scattering, the Berry curvature and side jump, although they can induce interesting transport phenomena under certain circumstances~\cite{Ma2023,Gong2024,PhysRevB.110.174423,PhysRevB.111.155127}.

The Boltzmann equation reproduced by the density matrix shows that there is no Pauli-blocking factors like $1 - f_{ {\bf k}^{\prime} }^{m} \left( t \right)$ in the collision integral for elastic scattering with static impurities [see Eqs.~(\ref{eq:J-sk}) and (\ref{eq:full-collision-integral})], in agreement with previous studies~\cite{Kohn-Luttinger1957,Luttinger1958,Sturman1984,Hankiewicz2006}. By contrast, many textbooks uniformly add the Pauli-blocking factors to the collision integral~\cite{Ziman1972, Ashcroft1976}, engendering spurious nonlinear terms when skew scattering is included. Actually, since our Hamiltonian $\hat{H}$ does not have electron-electron interaction, the resultant quantum Liouville equation~(\ref{eq:Liouville-1}) is linear in the density matrix $\hat{\rho}$, and there are no correlation effects, i.e., no $f_{ {\bf k} }^{m} \left( t \right)  f_{ {\bf k}^{\prime} }^{m} \left( t \right)$ terms. Although our derivation comes with a reduced single-particle density matrix, the result is the same if one operates with the full density matrix. This might seem counterintuitive, as one may expect that electron scattering processes from volume element $d{\bf k}$ about ${\bf k}$ to $d{\bf k}^{\prime}$ about ${\bf k}^{\prime}$ involve a fraction $f_{ {\bf k} }$ of states in $d{\bf k}$ that is occupied and a fraction $\left( 1 - f_{ {\bf k}^{\prime} } \right)$ in $d{\bf k}^{\prime}$ that is empty, and one is tempted to add $\left( 1 - f_{ {\bf k}^{\prime} } \right)$ to the collision integral to explicitly account for Pauli exclusion. This misconception can be resolved with the theory of scattering in quantum mechanics. Let us consider a general Hamiltonian $\hat{H} = \hat{H}_{0} + \hat{V}$, with $\vert l \rangle \equiv \vert m, {\bf k} \rangle$ being an eigenstate of $\hat{H}_{0}$, where $m$ is the band number, and ${\bf k}$ is the crystal momentum. According to the Lippmann-Schwinger equation, the eigenstate of $\hat{H}$, $\vert \psi_{l} \rangle$, satisfies $\vert \psi_{l} \rangle = \vert l \rangle + \frac{1}{ \varepsilon_{l} - \hat{H}_{0} + i \eta } \hat{V} \vert \psi_{l} \rangle$, where $\eta \to 0^{ + }$ and $ \hat{H}_{0} \vert l \rangle = \varepsilon_{l} \vert l \rangle $. Therefore,
\begin{equation}
\vert \psi_{l} \rangle = \vert l \rangle + \sum_{ l^{\prime} } \frac{1}{ \varepsilon_{l} - \varepsilon_{ l^{\prime} } + i \eta } \vert l^{\prime} \rangle \langle l^{\prime} \vert \hat{V} \vert l \rangle + \mathcal{O} \left( V^{2} \right).
\end{equation}
We see that the scattering state $\vert \psi_{l} \rangle$ is composed of an incident state $\vert l \rangle$ and scattered states $\vert l^{\prime} \rangle =  \vert m, {\bf k}^{\prime} \rangle$. Similar to the case for the mesoscopic Landauer-B{\"u}ttiker formalism~\cite{Datta1995}, there is no need for a transition from one state to another state, so there is no Pauli-blocking factor $1 - f_{ {\bf k}^{\prime} }^{m} \left( t \right)$ in the collision integral.

\section{Conclusion and Discussion}
The semiclassical Boltzmann equation is a handy and useful tool for transport problems. However, being semiclassical and borrowing heavily from the classical mechanics, the equation calls for close inspection from the perspective of quantum mechanics. Previous works tried to verify the Boltzmann equation with the Kubo formula and the Keldysh formalism. Alternatively, attempts have been made to verify the equation with the density matrix, but the issues of impurity scattering have not been fully resolved. We devised a systematic approach to directly compare the semiclassical Boltzmann equation with the quantum Liouville equation, reproducing the group velocity and the anomalous velocity generated by the Berry curvature. As for side jumps, we found that this quantum approach adds an additional longitudinal velocity to both the velocity and the distribution correction, which was not recognized in earlier works, and the semiclassical side-jump collision integral is an approximation of the quantum result at moderate temperatures. For skew scattering, we recover the scattering rate at orders $V^{3}$ and $V^{4}$ derived from the density matrix, reaching agreements with the semiclassical formalism and other methods. We show that the semiclassical Boltzmann equation is more than a mishmash of quantum mechanics and an analog of classical mechanics, instead it can trace its roots to quantum mechanics. 

Although we worked in the low-frequency limit in the derivation here and ignored interband transition, the density matrix formalism here is applicable to high-frequency scenarios, for example, nonlinear optics~\cite{Sipe2000}. We assume that impurity scattering is weak here, and the density of states is unaffected by disorder. In realistic systems, disorder effects can be strong, and the bands can be significantly renormalized by disorder~\cite{Groth2009}. It would also be interesting to generalize the treatment to nonlinear response, degenerate systems, magnetotransport and spatially dependent electric field in future works.

\begin{acknowledgments}
This work was financially supported by the National Natural Science Foundation of China (Grant No. 12350401), the National Key R\&D Program of China (Grant No. 2022YFA1403700, and No. 2024YFA1409003), and the Shanghai Science and Technology Innovation Action Plan (Grant No. 24LZ1400800).
\end{acknowledgments}

\bibliography{Bib}

\onecolumngrid
\newpage
\pagebreak
\widetext
\begin{center}
\textbf{\large Appendices}
\end{center}
\setcounter{equation}{0}
\setcounter{table}{0}
\setcounter{figure}{0}
\setcounter{section}{0}
\makeatletter 
\renewcommand{\thesubsection}{\Alph{subsection}}
\renewcommand{\thefigure}{S\arabic{figure}}
\renewcommand{\theequation}{S\arabic{equation}}
\renewcommand{\thetable}{S\arabic{table}}
\renewcommand{\bibnumfmt}[1]{[S#1]}
\renewcommand{\theHequation}{Supplement.\theequation}
\renewcommand{\theHfigure}{Supplement.\thefigure}
\renewcommand{\theHtable}{Supplement.\thetable}


\subsection{The position operator and the related commutator}
\label{sec:r-operator}
In this appendix, we derive the commutator of the position operator $\hat{ {\bf r} }$ and an operator $\hat{O}$ in the Bloch representation, $[ \hat{ {\bf r} }, \hat{O} ]$. For simplicity, in this appendix we work in the Schr{\"o}dinger picture and denote the Bloch basis in this picture as $\vert m, {\bf k} \rangle = e^{i {\bf k} \cdot {\bf r} } u_{m, {\bf k}} \left( {\bf r} \right)$, where $u_{m, {\bf k}} \left( {\bf r} \right)$ is the periodic part of the wave function.
\begin{align}
& \langle m, {\bf k} \vert [ \hat{ {\bf r} }, \hat{O} ] \vert m^{\prime}, {\bf k}^{\prime} \rangle \nonumber \\
=& \sum_{m^{\prime\prime}} \int d{\bf k}^{\prime\prime} \left( \langle m, {\bf k} \vert \hat{ {\bf r} } \vert m^{\prime\prime}, {\bf k}^{\prime\prime} \rangle \langle m^{\prime\prime}, {\bf k}^{\prime\prime} \vert \hat{O} \vert m^{\prime}, {\bf k}^{\prime} \rangle - \langle m, {\bf k} \vert \hat{O} \vert m^{\prime\prime}, {\bf k}^{\prime\prime} \rangle \langle m^{\prime\prime}, {\bf k}^{\prime\prime} \vert \hat{ {\bf r} } \vert m^{\prime}, {\bf k}^{\prime} \rangle \right) \nonumber \\
=& \sum_{m^{\prime\prime}} \int d{\bf k}^{\prime\prime} \int d{\bf r} \left\{ e^{- i {\bf k} \cdot {\bf r} } u^{*}_{m, {\bf k} } \left( {\bf r} \right) {\bf r} e^{i {\bf k}^{\prime\prime} \cdot {\bf r} } u_{ m^{\prime\prime}, {\bf k}^{\prime\prime} }\left( {\bf r} \right) O_{ {\bf k}^{\prime\prime} {\bf k}^{\prime} }^{ m^{\prime\prime} m^{\prime} } - O_{ {\bf k} {\bf k}^{\prime\prime} }^{ m m^{\prime\prime} } e^{- i {\bf k}^{\prime\prime} \cdot {\bf r} } u^{*}_{ m^{\prime\prime}, {\bf k}^{\prime\prime} } \left( {\bf r} \right) {\bf r} e^{ i {\bf k}^{\prime} \cdot {\bf r} } u_{m^{\prime}, {\bf k}^{\prime} } \left( {\bf r} \right) \right\} \nonumber \\
=& -i \sum_{m^{\prime\prime}} \int d{\bf k}^{\prime\prime} \int d{\bf r} \left\{ e^{- i {\bf k} \cdot {\bf r} } u^{*}_{m, {\bf k} } \left( {\bf r} \right) \left( \nabla_{ {\bf k}^{\prime\prime} } e^{i {\bf k}^{\prime\prime} \cdot {\bf r} } \right) u_{ m^{\prime\prime}, {\bf k}^{\prime\prime} }\left( {\bf r} \right) O_{ {\bf k}^{\prime\prime} {\bf k}^{\prime} }^{ m^{\prime\prime} m^{\prime} } + O_{ {\bf k} {\bf k}^{\prime\prime} }^{ m m^{\prime\prime} } \left( \nabla_{ {\bf k}^{\prime\prime} } e^{ - i {\bf k}^{\prime\prime} \cdot {\bf r} } \right) u^{*}_{ m^{\prime\prime}, {\bf k}^{\prime\prime} } \left( {\bf r} \right) e^{ i {\bf k}^{\prime} \cdot {\bf r} } u_{m^{\prime}, {\bf k}^{\prime} } \left( {\bf r} \right) \right\} \nonumber \\
=& i \sum_{m^{\prime\prime}} \int d{\bf k}^{\prime\prime} \int d{\bf r} \left\{ e^{- i {\bf k} \cdot {\bf r} } u^{*}_{m, {\bf k} } \left( {\bf r} \right)  e^{i {\bf k}^{\prime\prime} \cdot {\bf r} } \nabla_{ {\bf k}^{\prime\prime} } \left( u_{ m^{\prime\prime}, {\bf k}^{\prime\prime} }  \left( {\bf r} \right) O_{ {\bf k}^{\prime\prime} {\bf k}^{\prime} }^{ m^{\prime\prime} m^{\prime} } \right) + e^{ - i {\bf k}^{\prime\prime} \cdot {\bf r} } \nabla_{ {\bf k}^{\prime\prime} } \left( O_{ {\bf k} {\bf k}^{\prime\prime} }^{ m m^{\prime\prime} } u^{*}_{ m^{\prime\prime}, {\bf k}^{\prime\prime} } \left( {\bf r} \right)  \right) e^{ i {\bf k}^{\prime} \cdot {\bf r} } u_{m^{\prime}, {\bf k}^{\prime} } \left( {\bf r} \right) \right\} \nonumber \\
=& i \sum_{m^{\prime\prime}} \int d{\bf k}^{\prime\prime} \int d{\bf r} \bigg\{ e^{- i \left( {\bf k} - {\bf k}^{\prime\prime} \right) \cdot {\bf r} } u^{*}_{m, {\bf k} } \left( {\bf r} \right)   O_{ {\bf k}^{\prime\prime} {\bf k}^{\prime} }^{ m^{\prime\prime} m^{\prime} } \nabla_{ {\bf k}^{\prime\prime} } u_{ m^{\prime\prime}, {\bf k}^{\prime\prime} } \left( {\bf r} \right)   + e^{- i \left( {\bf k} - {\bf k}^{\prime\prime} \right) \cdot {\bf r} } u^{*}_{m, {\bf k} } \left( {\bf r} \right) u_{ m^{\prime\prime}, {\bf k}^{\prime\prime} } \left( {\bf r} \right) \nabla_{ {\bf k}^{\prime\prime} } O_{ {\bf k}^{\prime\prime} {\bf k}^{\prime} }^{ m^{\prime\prime} m^{\prime} } \nonumber \\
& + e^{ - i \left( {\bf k}^{\prime\prime} - {\bf k}^{\prime} \right) \cdot {\bf r} } O_{ {\bf k} {\bf k}^{\prime\prime} }^{ m m^{\prime\prime} } u_{m^{\prime}, {\bf k}^{\prime} } \left( {\bf r} \right) \nabla_{ {\bf k}^{\prime\prime} } u^{*}_{ m^{\prime\prime}, {\bf k}^{\prime\prime} } \left( {\bf r} \right)  + e^{ - i \left( {\bf k}^{\prime\prime} - {\bf k}^{\prime} \right) \cdot {\bf r} } u^{*}_{ m^{\prime\prime}, {\bf k}^{\prime\prime} } \left( {\bf r} \right) u_{m^{\prime}, {\bf k}^{\prime} } \left( {\bf r} \right) \nabla_{ {\bf k}^{\prime\prime} } O_{ {\bf k} {\bf k}^{\prime\prime} }^{ m m^{\prime\prime} } \bigg\} \nonumber \\
=& i \sum_{m^{\prime\prime}} \int d{\bf k}^{\prime\prime} \int d{\bf r} \bigg\{ \delta \left( {\bf k} - {\bf k}^{\prime\prime} \right) u^{*}_{m, {\bf k} } \left( {\bf r} \right)   O_{ {\bf k}^{\prime\prime} {\bf k}^{\prime} }^{ m^{\prime\prime} m^{\prime} } \nabla_{ {\bf k}^{\prime\prime} } u_{ m^{\prime\prime}, {\bf k}^{\prime\prime} } \left( {\bf r} \right) + \delta \left( {\bf k}^{\prime\prime} - {\bf k}^{\prime} \right) O_{ {\bf k} {\bf k}^{\prime\prime} }^{ m m^{\prime\prime} } u_{m^{\prime}, {\bf k}^{\prime} } \left( {\bf r} \right) \nabla_{ {\bf k}^{\prime\prime} } u^{*}_{ m^{\prime\prime}, {\bf k}^{\prime\prime} } \left( {\bf r} \right) \bigg\} \nonumber \\
& + i \sum_{m^{\prime\prime}} \int d{\bf k}^{\prime\prime} \bigg\{ \delta \left( {\bf k} - {\bf k}^{\prime\prime} \right) \delta_{m, m^{\prime\prime} } \nabla_{ {\bf k}^{\prime\prime} } O_{ {\bf k}^{\prime\prime} {\bf k}^{\prime} }^{ m^{\prime\prime} m^{\prime} }  + \delta \left( {\bf k}^{\prime\prime} - {\bf k}^{\prime} \right) \delta_{ m^{\prime\prime} m^{\prime} } \nabla_{ {\bf k}^{\prime\prime} } O_{ {\bf k} {\bf k}^{\prime\prime} }^{ m m^{\prime\prime} } \bigg\} \nonumber \\
=& \sum_{m^{\prime\prime}} \left( {\bf A}_{ {\bf k} }^{m m^{\prime\prime} } O_{ {\bf k} {\bf k}^{\prime} }^{ m^{\prime\prime} m^{\prime} } - O_{ {\bf k} {\bf k}^{\prime} }^{ m m^{\prime\prime} } {\bf A}_{ {\bf k}^{\prime} }^{m^{\prime\prime} m^{\prime} } \right) + i \left( \nabla_{ {\bf k} } + \nabla_{ {\bf k}^{\prime} } \right) O_{ {\bf k} {\bf k}^{\prime} }^{ m m^{\prime} },
\end{align}
where we have taken into account the face that $e^{- i \left( {\bf k} - {\bf k}^{\prime} \right) \cdot {\bf r} } u^{*}_{m, {\bf k} } \left( {\bf r} \right) \nabla_{ {\bf k}^{\prime} } u_{ m^{\prime}, {\bf k}^{\prime} } \left( {\bf r} \right)$ is a periodic function of ${\bf r}$~\cite{Blount1962,Callaway1974}, ${\bf A}_{ {\bf k} }^{m m^{\prime} } = i \int d{\bf r} u^{*}_{m, {\bf k} } \left( {\bf r} \right) \nabla_{ {\bf k} } u_{ m^{\prime}, {\bf k} } \left( {\bf r} \right)$ is the interband Berry connection, and the shorthand notation $O_{ {\bf k} {\bf k}^{\prime} }^{ m m^{\prime} } = \langle m, {\bf k} \vert \hat{O} \vert m^{\prime}, {\bf k}^{\prime} \rangle$ is introduced.

As a special case, if $\hat{O}$ commutes with $\hat{ {\bf r} }$, we have
\begin{align}
\sum_{m^{\prime\prime}} \left( {\bf A}_{ {\bf k} }^{m m^{\prime\prime} } O_{ {\bf k} {\bf k}^{\prime} }^{ m^{\prime\prime} m^{\prime} } - O_{ {\bf k} {\bf k}^{\prime} }^{ m m^{\prime\prime} } {\bf A}_{ {\bf k}^{\prime} }^{m^{\prime\prime} m^{\prime} } \right) = - i \left( \nabla_{ {\bf k} } + \nabla_{ {\bf k}^{\prime} } \right) O_{ {\bf k} {\bf k}^{\prime} }^{ m m^{\prime} }.
\end{align}
This happens when $\hat{O}$ can be expressed as a function of ${\bf r}$. The equation above is utilized in the main text to get the side-jump position shift.

As for the commutator of the position operator and the density matrix at equilibrium $[ \hat{ {\bf r} }, \hat{\rho}_{0} ]$,
\begin{align}
& \langle m, {\bf k} \vert [ \hat{ {\bf r} }, \hat{\rho}_{0} ] \vert m^{\prime}, {\bf k}^{\prime} \rangle \nonumber \\
=& \langle m, {\bf k} \vert \hat{ {\bf r} } \vert m^{\prime}, {\bf k}^{\prime} \rangle \left( f_{0, {\bf k}^{\prime} }^{m^{\prime} } - f_{0, {\bf k} }^{m} \right)  \nonumber \\
=& \int d{\bf r} e^{- i {\bf k} \cdot {\bf r} } u^{*}_{m, {\bf k} } \left( {\bf r} \right) {\bf r} e^{i {\bf k}^{\prime} \cdot {\bf r} } u_{ m^{\prime}, {\bf k}^{\prime} }\left( {\bf r} \right) \left( f_{0, {\bf k}^{\prime} }^{m^{\prime} } - f_{0, {\bf k} }^{m} \right) \nonumber \\
=& - i \int d{\bf r} e^{- i {\bf k} \cdot {\bf r} } u^{*}_{m, {\bf k} } \left( {\bf r} \right) \left( \nabla_{ {\bf k}^{\prime} } e^{i {\bf k}^{\prime} \cdot {\bf r} } \right) u_{ m^{\prime}, {\bf k}^{\prime} }\left( {\bf r} \right) \left( f_{0, {\bf k}^{\prime} }^{m^{\prime} } - f_{0, {\bf k} }^{m} \right) \nonumber \\
=& i \int d{\bf r} e^{- i {\bf k} \cdot {\bf r} } u^{*}_{m, {\bf k} } \left( {\bf r} \right) e^{i {\bf k}^{\prime} \cdot {\bf r} } u_{ m^{\prime}, {\bf k}^{\prime} }\left( {\bf r} \right)  \nabla_{ {\bf k}^{\prime} } f_{0, {\bf k}^{\prime} }^{m^{\prime} } + i \int d{\bf r} e^{- i {\bf k} \cdot {\bf r} } u^{*}_{m, {\bf k} } \left( {\bf r} \right) e^{i {\bf k}^{\prime} \cdot {\bf r} } \left[ \nabla_{ {\bf k}^{\prime} } u_{ m^{\prime}, {\bf k}^{\prime} }\left( {\bf r} \right) \right]  \left( f_{0, {\bf k}^{\prime} }^{m^{\prime} } - f_{0, {\bf k} }^{m} \right) \nonumber \\
=& i \delta_{ m m^{\prime} } \delta \left( {\bf k} - {\bf k}^{\prime} \right) \nabla_{ {\bf k}^{\prime} } f_{0, {\bf k}^{\prime} }^{m^{\prime} } + \delta \left( {\bf k} - {\bf k}^{\prime} \right) {\bf A}_{ {\bf k} }^{m m^{\prime} } \left( f_{0, {\bf k}^{\prime} }^{m^{\prime} } - f_{0, {\bf k} }^{m} \right),
\end{align}
where $f_{0, {\bf k} }^{m} = \langle m, {\bf k} \vert \hat{\rho}_{0} \vert m, {\bf k} \rangle$ is the equilibrium distribution of band $m$ and momentum ${\bf k}$.

\subsection{Disorder averaging}
\label{sec:disorder-averaging}
We follow the widely used procedure to handle impurity scattering~\cite{Kohn-Luttinger1957,Doniach1998,Culcer2017,Atencia2022}. We assume the impurities are randomly distributed and uncorrelated. In practice, we take the average of all possible impurity configurations with a fixed impurity density. To facilitate evaluation, we take the Fourier transform of the impurity potential, $\mathcal{U} \left( {\bf q} \right) \equiv \int \frac{ d {\bf r} }{ \Omega } e^{ - i {\bf q} \cdot {\bf r} } U \left( {\bf r} \right)$, where $U \left( {\bf r} \right) = \sum_{i} V \left( {\bf r} - {\bf R}_{i} \right)$, $V \left( {\bf r} \right)$ is the single-impurity potential, and $\Omega$ here is the volume of the material. Obviously,
\begin{align}
\mathcal{U} \left( {\bf q} \right) = & \sum_{i} \int \frac{ d {\bf r} }{ \Omega } e^{ - i {\bf q} \cdot {\bf r} } V \left( {\bf r} - {\bf R}_{i} \right) \nonumber \\
= & \sum_{i} \sum_{ {\bf q}^{\prime} } \int \frac{ d {\bf r} }{ \Omega } e^{ - i {\bf q} \cdot {\bf r} + i {\bf q}^{\prime} \cdot \left( {\bf r} - {\bf R}_{i} \right) } \mathcal{V} \left( {\bf q}^{\prime} \right) \nonumber \\
= & \sum_{i} e^{ - i {\bf q} \cdot {\bf R}_{i}  } \mathcal{V} \left( {\bf q} \right) ,
\end{align}
where $\mathcal{V} \left( {\bf q} \right) \equiv \int \frac{ d {\bf r} }{ \Omega } e^{ - i {\bf q} \cdot {\bf r} } V \left( {\bf r} \right)$. As a special case, for the short-range impurity potential $V \left( {\bf r} \right) = V_{0} \delta \left( {\bf r} \right)$, the Fourier transform reads $\mathcal{V} \left( {\bf q} \right) = \frac{ V_{ 0 } }{ \Omega }$. The ensemble average of $\mathcal{U} \left( {\bf q} \right)$ is calculated as follows:
\begin{align}
\label{eq:S-single-U-average}
\overline{ \mathcal{U} \left( {\bf q} \right)  } = & \sum_{i} \int \frac{ d {\bf R}_{1} }{ \Omega } \int \frac{ d {\bf R}_{2} }{ \Omega } \cdots \int \frac{ d {\bf R}_{N} }{ \Omega } e^{ - i {\bf q} \cdot {\bf R}_{i}  } \mathcal{V} \left( {\bf q} \right) \nonumber \\
= & \sum_{i} \int \frac{ d {\bf R}_{i} }{ \Omega } e^{ - i {\bf q} \cdot {\bf R}_{i}  } \mathcal{V} \left( {\bf q} \right) \nonumber \\
= & N \mathcal{V} \left( 0 \right) ,
\end{align}
where $N$ is the total number of impurities. $\overline{ \mathcal{U} \left( {\bf q} \right)}$ is independent of ${\bf q}$, which amounts to a shift in band energy, and it is usually absorbed into band dispersion. As a result, $\overline{ \mathcal{U} \left( {\bf q} \right)}$ is discarded~\cite{Kohn-Luttinger1957,Culcer2017}.

The conventional scattering involves $\langle U_{ {\bf k} {\bf k}^{ \prime } }^{ m m^{ \prime } } U_{ {\bf k}^{ \prime} {\bf k} }^{ m^{ \prime \prime } m^{ \prime \prime \prime } } \rangle$, where $\langle \cdots \rangle$ stands for the impurity configuration ensemble average, i.e., $\overline{ \mathcal{U} \left( {\bf k} - {\bf k}^{ \prime } \right) \mathcal{U} \left( {\bf k}^{ \prime } - {\bf k} \right) }$.
\begin{align}
\label{eq:S-two-U-average}
\overline{ \mathcal{U} \left( {\bf k} - {\bf k}^{ \prime } \right) \mathcal{U} \left( {\bf k}^{ \prime } - {\bf k} \right) } = & \sum_{i, j} \int \frac{ d {\bf R}_{1} }{ \Omega } \int \frac{ d {\bf R}_{2} }{ \Omega } \cdots \int \frac{ d {\bf R}_{N} }{ \Omega } e^{ - i \left( {\bf k} - {\bf k}^{ \prime } \right) \cdot {\bf R}_{i}  } \mathcal{V} \left( {\bf k} - {\bf k}^{ \prime } \right) e^{ - i \left( {\bf k}^{ \prime } - {\bf k} \right) \cdot {\bf R}_{j}  } \mathcal{V} \left( {\bf k}^{ \prime } - {\bf k} \right) \nonumber \\
= & \sum_{i} \int \frac{ d {\bf R}_{i} }{ \Omega } \mathcal{V} \left( {\bf k} - {\bf k}^{ \prime } \right) \mathcal{V} \left( {\bf k}^{ \prime } - {\bf k} \right) + \sum_{i \ne j} \int \frac{ d {\bf R}_{i} }{ \Omega } \int \frac{ d {\bf R}_{j} }{ \Omega } e^{ - i \left( {\bf k} - {\bf k}^{ \prime } \right) \cdot \left( {\bf R}_{i} - {\bf R}_{j} \right) } \mathcal{V} \left( {\bf k} - {\bf k}^{ \prime } \right) \mathcal{V} \left( {\bf k}^{ \prime } - {\bf k} \right) \nonumber \\
= & N \mathcal{V} \left( {\bf k} - {\bf k}^{ \prime } \right) \mathcal{V} \left( {\bf k}^{ \prime } - {\bf k} \right) + N \left( N - 1 \right) \delta_{ {\bf k}, {\bf k}^{ \prime } } \left[ \mathcal{V} \left( 0 \right) \right]^{2} \nonumber \\
\approx & N \vert \mathcal{V} \left( {\bf k} - {\bf k}^{ \prime } \right) \vert^{2},
\end{align}
where $\delta_{ {\bf k}, {\bf k}^{ \prime } }$ is the Kronecker delta. We dropped $N \left( N - 1 \right) \delta_{ {\bf k}, {\bf k}^{ \prime } } \left[ \mathcal{V} \left( 0 \right) \right]^{2}$ because it is trivial and does not contribute to the scattering. We also utilized $\mathcal{V} \left( {\bf q} \right) = \mathcal{V}^{*} \left( - {\bf q} \right)$, since $V \left( {\bf r} \right)$ is real. According to Eq.~(\ref{eq:S-two-U-average}), 
\begin{align}
\langle U_{ {\bf k} {\bf k}^{ \prime } }^{ m m^{ \prime } } U_{ {\bf k}^{ \prime} {\bf k} }^{ m^{ \prime \prime } m^{ \prime \prime \prime } } \rangle = n_{i} V_{ {\bf k} {\bf k}^{ \prime } }^{ m m^{ \prime } } V_{ {\bf k}^{ \prime} {\bf k} }^{ m^{ \prime \prime } m^{ \prime \prime \prime } },
\end{align}
where $V_{ {\bf k} {\bf k}^{ \prime } }^{ m m^{ \prime } } \equiv \langle m, {\bf k}, t \vert \hat{V}_{\mathrm{I}}(t) \vert m^{\prime}, {\bf k}^{ \prime }, t \rangle$. As we follow common practices and take $\Omega$ to be the unit volume, the total number of impurities $N$ is replaced by the impurity density $n_{i}$.

The skew-scattering rate at order $V^{3}$ contains $\mathrm{Im} \left( \langle U_{ {\bf k} {\bf k}^{ \prime } }^{ m m } U_{ {\bf k}^{ \prime} {\bf k}^{ \prime \prime } }^{ m m } U_{ {\bf k}^{ \prime \prime } {\bf k} }^{ m m } \rangle \right)$. Here, we consider the corresponding ensemble average $\overline{ \mathcal{U} \left( {\bf k} - {\bf k}^{ \prime } \right) \mathcal{U} \left( {\bf k}^{ \prime } - {\bf k}^{ \prime \prime } \right) \mathcal{U} \left( {\bf k}^{ \prime \prime } - {\bf k} \right) }$,
\begin{align}
\label{eq:S-three-U-average}
\overline{ \mathcal{U} \left( {\bf k} - {\bf k}^{ \prime } \right) \mathcal{U} \left( {\bf k}^{ \prime } - {\bf k}^{ \prime \prime } \right) \mathcal{U} \left( {\bf k}^{ \prime \prime } - {\bf k} \right) } = & \sum_{i, j, l } \int \frac{ d {\bf R}_{1} }{ \Omega } \int \frac{ d {\bf R}_{2} }{ \Omega } \cdots \int \frac{ d {\bf R}_{N} }{ \Omega } e^{ - i \left( {\bf k} - {\bf k}^{ \prime } \right) \cdot {\bf R}_{i} } \mathcal{V} \left( {\bf k} - {\bf k}^{ \prime } \right) e^{ - i \left( {\bf k}^{ \prime } - {\bf k}^{ \prime \prime } \right) \cdot {\bf R}_{j} } \mathcal{V} \left( {\bf k}^{ \prime } - {\bf k}^{ \prime \prime } \right) \nonumber \\
& e^{ - i \left( {\bf k}^{ \prime \prime } - {\bf k} \right) \cdot {\bf R}_{l} } \mathcal{V} \left( {\bf k}^{ \prime \prime } - {\bf k} \right) \nonumber \\
= & \bigg\{ \sum_{ i } \int \frac{ d {\bf R}_{i} }{ \Omega } + \sum_{i \ne l } \int \frac{ d {\bf R}_{i} }{ \Omega } \int \frac{ d {\bf R}_{l} }{ \Omega }  e^{ - i \left( {\bf k} - {\bf k}^{ \prime \prime } \right) \cdot \left( {\bf R}_{i} - {\bf R}_{l} \right) } \nonumber \\
& + \sum_{i \ne j } \int \frac{ d {\bf R}_{i} }{ \Omega } \int \frac{ d {\bf R}_{j} }{ \Omega } e^{ - i \left( {\bf k}^{ \prime \prime } - {\bf k}^{ \prime } \right) \cdot \left( {\bf R}_{i} - {\bf R}_{l} \right) } + \sum_{ i \ne j } \int \frac{ d {\bf R}_{i} }{ \Omega } \int \frac{ d {\bf R}_{j} }{ \Omega }  e^{ - i \left( {\bf k} - {\bf k}^{ \prime } \right) \cdot \left( {\bf R}_{i} - {\bf R}_{j} \right) } \nonumber \\
& + \sum_{ i } \sum_{ j \ne i } \sum_{ \substack{ l \ne j \\ l \ne i } } \int \frac{ d {\bf R}_{i} }{ \Omega } \int \frac{ d {\bf R}_{j} }{ \Omega } \int \frac{ d {\bf R}_{l} }{ \Omega } e^{ - i \left( {\bf k} - {\bf k}^{ \prime } \right) \cdot {\bf R}_{i} - i \left( {\bf k}^{ \prime } - {\bf k}^{ \prime \prime } \right) \cdot {\bf R}_{j}  - i \left( {\bf k}^{ \prime \prime } - {\bf k} \right) \cdot {\bf R}_{l} } \bigg\} \nonumber \\
& \mathcal{V} \left( {\bf k} - {\bf k}^{ \prime } \right) \mathcal{V} \left( {\bf k}^{ \prime } - {\bf k}^{ \prime \prime } \right) \mathcal{V} \left( {\bf k}^{ \prime \prime } - {\bf k} \right) \nonumber \\
= & N \mathcal{V} \left( {\bf k} - {\bf k}^{ \prime } \right) \mathcal{V} \left( {\bf k}^{ \prime } - {\bf k}^{ \prime \prime } \right) \mathcal{V} \left( {\bf k}^{ \prime \prime } - {\bf k} \right) + N \left( N - 1 \right) \delta_{ {\bf k}, {\bf k}^{ \prime \prime } } \mathcal{V} \left( 0 \right) \vert \mathcal{V} \left( {\bf k} - {\bf k}^{ \prime } \right) \vert^{2} \nonumber \\
& + N \left( N - 1 \right) \delta_{ {\bf k}^{ \prime \prime }, {\bf k}^{ \prime } } \mathcal{V} \left( 0 \right) \vert \mathcal{V} \left( {\bf k} - {\bf k}^{ \prime } \right) \vert^{2} + N \left( N - 1 \right) \delta_{ {\bf k}, {\bf k}^{ \prime } } \mathcal{V} \left( 0 \right) \vert \mathcal{V} \left( {\bf k}^{ \prime \prime } - {\bf k} \right) \vert^{2} \nonumber \\
& + N \left( N - 1 \right) \left( N - 2 \right) \delta_{ {\bf k}, {\bf k}^{ \prime } } \delta_{ {\bf k}, {\bf k}^{ \prime \prime } } \left[ \mathcal{V} \left( 0 \right) \right]^{3} \nonumber \\
\approx & N \mathcal{V} \left( {\bf k} - {\bf k}^{ \prime } \right) \mathcal{V} \left( {\bf k}^{ \prime } - {\bf k}^{ \prime \prime } \right) \mathcal{V} \left( {\bf k}^{ \prime \prime } - {\bf k} \right) ,
\end{align}
where we have kept only terms contributing to skew-scattering rates. Similarly,
\begin{align}
\langle U_{ {\bf k} {\bf k}^{ \prime } }^{ m m } U_{ {\bf k}^{ \prime} {\bf k}^{ \prime \prime } }^{ m m } U_{ {\bf k}^{ \prime \prime } {\bf k} }^{ m m } \rangle = n_{i} V_{ {\bf k} {\bf k}^{ \prime } }^{ m m } V_{ {\bf k}^{ \prime} {\bf k}^{ \prime \prime } }^{ m m } V_{ {\bf k}^{ \prime \prime } {\bf k} }^{ m m }.
\end{align}

As for Gaussian skew scattering, we consider the impurity configuration ensemble average for $U_{ {\bf k}_{1} {\bf k}_{2} }^{ m_{1} m_{2} } U_{ {\bf k}_{2} {\bf k}_{3} }^{ m_{2} m_{3} } U_{ {\bf k}_{3} {\bf k}_{4} }^{ m_{3} m_{4} } U_{ {\bf k}_{4} {\bf k}_{1} }^{ m_{4} m_{1} }$, namely, $\overline{ \mathcal{U} \left( {\bf k}_{1} - {\bf k}_{2} \right) \mathcal{U} \left( {\bf k}_{2} - {\bf k}_{3} \right) \mathcal{U} \left( {\bf k}_{3} - {\bf k}_{4} \right) \mathcal{U} \left( {\bf k}_{4} - {\bf k}_{1} \right) }$. In the same fashion,
\begin{align}
& \overline{ \mathcal{U} \left( {\bf k}_{1} - {\bf k}_{2} \right) \mathcal{U} \left( {\bf k}_{2} - {\bf k}_{3} \right) \mathcal{U} \left( {\bf k}_{3} - {\bf k}_{4} \right) \mathcal{U} \left( {\bf k}_{4} - {\bf k}_{1} \right) } \nonumber \\
\approx & \bigg[ N + N \left( N - 1 \right) \left( \delta_{ {\bf k}_{1} + {\bf k}_{3}, {\bf k}_{2} + {\bf k}_{4} } + \delta_{ {\bf k}_{1}, {\bf k}_{3} } + \delta_{ {\bf k}_{2}, {\bf k}_{4} } \right) \bigg] \mathcal{V} \left( {\bf k}_{1} - {\bf k}_{2} \right) \mathcal{V} \left( {\bf k}_{2} - {\bf k}_{3} \right) \mathcal{V} \left( {\bf k}_{3} - {\bf k}_{4} \right) \mathcal{V} \left( {\bf k}_{4} - {\bf k}_{1} \right)\nonumber \\
\approx & \bigg[ N + N^{ 2 } \left( \delta_{ {\bf k}_{1} + {\bf k}_{3}, {\bf k}_{2} + {\bf k}_{4} } + \delta_{ {\bf k}_{1}, {\bf k}_{3} } + \delta_{ {\bf k}_{2}, {\bf k}_{4} } \right) \bigg] \mathcal{V} \left( {\bf k}_{1} - {\bf k}_{2} \right) \mathcal{V} \left( {\bf k}_{2} - {\bf k}_{3} \right) \mathcal{V} \left( {\bf k}_{3} - {\bf k}_{4} \right) \mathcal{V} \left( {\bf k}_{4} - {\bf k}_{1} \right).
\end{align}
The term proportional to $N^{ 2 }$ leads to Gaussian skew scattering, and the scaling of the relevant current is the same as the current induced by the side jump ~\cite{Sinitsyn2006,Sinitsyn2007PRB,Nagaosa2010}. The classic paper by Kohn and Luttinger~\cite{Kohn-Luttinger1957} only included $\delta_{ {\bf k}_{1} + {\bf k}_{3}, {\bf k}_{2} + {\bf k}_{4} }$ in the parentheses, which corresponds to crossing diagrams in the Kubo-Streda formalism~\cite{Sinitsyn2007PRB} and neglecting the other two terms. In recent years, studies employing the Boltzmann equation frequently take the non-crossing approximation (NCA) and keep only the two terms, $ \delta_{ {\bf k}_{1}, {\bf k}_{3} }$ and $\delta_{ {\bf k}_{2}, {\bf k}_{4} }$, in the parentheses~\cite{Sinitsyn2007PRB}. We keep both the crossing and the non-crossing parts,
\begin{align}
\langle U_{ {\bf k}_{1} {\bf k}_{2} }^{ m_{1} m_{2} } U_{ {\bf k}_{2} {\bf k}_{3} }^{ m_{2} m_{3} } U_{ {\bf k}_{3} {\bf k}_{4} }^{ m_{3} m_{4} } U_{ {\bf k}_{4} {\bf k}_{1} }^{ m_{4} m_{1} } \rangle = n_{i}^{2} \left( \delta_{ {\bf k}_{1} + {\bf k}_{3}, {\bf k}_{2} + {\bf k}_{4} } + \delta_{ {\bf k}_{1}, {\bf k}_{3} } + \delta_{ {\bf k}_{2}, {\bf k}_{4} } \right) V_{ {\bf k}_{1} {\bf k}_{2} }^{ m_{1} m_{2} } V_{ {\bf k}_{2} {\bf k}_{3} }^{ m_{2} m_{3} } V_{ {\bf k}_{3} {\bf k}_{4} }^{ m_{3} m_{4} } V_{ {\bf k}_{4} {\bf k}_{1} }^{ m_{4} m_{1} }.
\end{align}

\end{document}